\documentclass[]{jot}

\usepackage{multirow}
\usepackage{graphicx}
\usepackage{latexsym}
\usepackage{amsmath}
\usepackage{stmaryrd}

\usepackage{amssymb}
\usepackage{enumitem}
\usepackage{listings}
\usepackage{color}
\usepackage{array}
\usepackage{tikz}
\usepackage{multirow}
\usetikzlibrary{arrows}
\usetikzlibrary{automata}
\usetikzlibrary{positioning}
\usepackage{soul}
\usepackage{syntax}
\usepackage{todonotes}
\usepackage{subcaption}
\usepackage[altpo]{backnaur}
\usepackage{hyperref}
\usepackage{float}
\usepackage{fancyvrb}
\usepackage{cancel}
\usepackage{relsize}
\usepackage{pifont}

\newtheorem{definition}{Definition}

\newtheorem{example}{Example}[section]
\floatstyle{ruled}
\newfloat{grammar}{htbp}{lop}
\floatname{grammar}{Grammar}

\articletype{regular} 

\title{PICKLES: a Natural Language Framework for Requirement Specification and Model-Based Testing}

\author{Mar\'ia Bel\'en Rodr\'iguez}
\author{Petra van den Bos}

\affil{Formal Methods \& Tools, University of Twente, The Netherlands}


\runningtitle{PICKLES: a Natural Language Framework for Requirement Specification and Model-Based Testing} 

\definecolor{dkblue}{rgb}{0.0,0.0,0.9}
\definecolor{dkgreen}{rgb}{0,0.55,0}
\definecolor{dkred}{rgb}{0.6,0,0}
\definecolor{goldenrod}{rgb}{0.8,0.6,0}
\definecolor{purple}{rgb}{0.5,0,0.5}
\definecolor{teal}{rgb}{0,0.5,0.5}
\definecolor{gray}{rgb}{0.4,0.4,0.4}
\definecolor{black}{rgb}{0,0,0}

\lstdefinelanguage{PicklesDSL}{
    basicstyle=\scriptsize\ttfamily,
    backgroundcolor=\color{white},
    morekeywords=[1]{Scenario,Variable,Settings,Given,When,Then,And,such,that,exactly,not,at,most,least,elements,where,each,element,
     has, a, value, is, variable, of, type, with, range, an, attributes},
    keywordstyle=[1]\color{dkblue},
    morekeywords=[2]{equal,not,between,and,to,exactly,than,lower,greater,or,at,least},
    keywordstyle=[2]\color{dkred},
    morekeywords=[3]{AND,OR},
    keywordstyle=[3]\color{purple},
    morekeywords=[4]{string,boolean,decimal,integer,array,structure},
    keywordstyle=[4]\color{goldenrod},
    morestring=[b]",
    stringstyle=\color{dkgreen},
}

\lstdefinelanguage{Gherkin}{
    basicstyle=\scriptsize\ttfamily,
    backgroundcolor=\color{white},
    morekeywords=[1]{Given,When,Then,And},
    keywordstyle=[1]\color{dkblue},
}

\tikzset{
  smalllabel/.style={font=\scriptsize}
}
\tikzset{
  dashdot/.style = {dash pattern=on 3pt off 2pt on 0.5pt off 2pt}
}

\newcommand{\sortfun}[1]{\mathsf{type}_v(#1)}
\newcommand{\sortufun}[1]{\mathsf{type}_u(#1)}
\newcommand{\domfun}[1]{\mathsf{dom}(#1)}

\newcommand{\getValue}[2][]{\mathsf{get}_{#1}(#2)}

\newcommand{\semmap}[2][]{\llbracket #2 \rrbracket_\mathsf{#1}}
\newcommand{\invsemmap}[2][]{\llbracket #2 \rrbracket_\mathsf{#1}^{-1}}

\newcommand{\word}[2][]{%
  \if\relax\detokenize{#1}\relax
    w_{\text{#2}}%
  \else
    w_{\text{#2}_{#1}}%
  \fi
}

\newcommand{\STS}{\mathcal{S}}
\newcommand{\nonterm}[1]{\mathsf{#1}}

\newcommand{\disj}{\curlywedge}
\newcommand{\bigdisj}{\mathop{\raisebox{0.1ex}{\Large$\curlywedge$}}}
\newcommand{\seqcomp}{\triangleright}

\newcommand{\locset}{\mathcal{L}}

\newcommand{\switchrel}{\mathcal{R}}
\newcommand{\vlset}{\mathcal{V}_{l}}
\newcommand{\vpset}{\mathcal{V}_{p}}
\newcommand{\interact}{\mathcal{I}}
\newcommand{\Terms}[2][]{\mathcal{T}_#1(#2)}

\newcommand{\STSset}{\frak{S}}

\newcommand{\term}{\mathcal{T}}
\newcommand{\boolterm}{\mathcal{T}_{Bool}}
\newcommand{\valueuniv}{\mathcal{U}}

\newcommand{\true}{\mathit{true}}
\newcommand{\false}{\mathit{false}}

\newcommand{\ini}{\mathit{ini}}

\newenvironment{smallmath} {\begingroup\small\[\aligned} {\endaligned\]\endgroup}

\newcommand{\composition}{
\begin{tikzpicture}[->, >=stealth, auto, node distance=3cm, thick,
    every state/.style={circle, draw, minimum size=1.2cm, thick, font=\huge},
    every node/.append style={font=\LARGE}, 
    initial text=, scale=0.52, transform shape]

\node[state, initial, initial where=above] (l0) {$l_0$};

\node[state, left=2.5cm of l0] (l3) {$l_3$};
\node[state, below=1.5cm of l0] (l4) {$l_4$};
\node[state, right=2.5cm of l0] (l5) {$l_5$};

\node[state, above=2.5cm of l5] (l1) {$l_1$};
\node[state, right=2.5cm of l1] (l2) {$l_2$};

\node[state, left=2.75cm of l3] (l6) {$l_6$};
\node[state, below=1.5cm of l4] (l7) {$l_7$};
\node[state, right=2.75cm of l5] (l8) {$l_8$};

\node[state, below left=of l6] (l9) {$l_9$};
\node[state, below right=of l8] (l10) {$l_{10}$};

\node[state, below left=2.7cm of l7] (l12) {$l_{12}$};
\node[state, left=4.5cm of l12] (l11) {$l_{11}$};
\node[state, below right=2.7cm of l7] (l13) {$l_{13}$};
\node[state, right=4.5cm of l13] (l14) {$l_{14}$};

\node[state, below=1.5cm of l9] (l15) {$l_{15}$};
\node[state, below=1.5cm of l10] (l16) {$l_{16}$};
\node[state, below=2.5cm of l11] (l17) {$l_{17}$};
\node[state, below=2.5cm of l12] (l18) {$l_{18}$};
\node[state, below=2.5cm of l13] (l19) {$l_{19}$};
\node[state, below=2.5cm of l14] (l20) {$l_{20}$};

\path
    (l0) edge[dashdot] node[above, yshift=0.2cm, xshift=-0.5cm] {\shortstack[l]{$i_2$?$\phi_i^4$}} (l1)
    (l1) edge[dashdot] node[above, yshift=0.10cm] {\shortstack[l]{$o_2$!$\phi_o^4$}} (l2)

    (l0) edge[dashed] node[right=0.2cm] {\shortstack[l]{$i_1$?$\phi_i^1$}} (l4)
    (l4) edge[dashed] node[right=0.2cm] {\shortstack[l]{$o_1$!$\phi_o^1$}} (l7)
    (l7) edge[dashed] node[above, yshift=0.25cm] {\shortstack[l]{$i_1$?$\phi_i^1$}} (l11)
    (l11) edge[dashed] node[right] {\shortstack[l]{$o_1$!$\phi_o^1$}} (l17)
    (l17) edge[dashed, bend left] node[left] {\shortstack[l]{$i_1$?$\phi_i^1$}} (l11)

    (l0) edge[dotted] node[above, yshift=0.10cm] {\shortstack[l]{$i_1$?$\phi_i^2$}} (l3)
    (l3) edge[dotted] node[above, yshift=0.10cm] {\shortstack[l]{$o_1$!$\phi_o^2$}} (l6)
    (l6) edge[dashdot] node[left, yshift=0.2cm] {\shortstack[l]{$i_2$?$\phi_i^4$}} (l9)
    (l9) edge[dashdot] node[left] {\shortstack[l]{$o_2$!$\phi_o^4$}} (l15)
    
    (l7) edge[dotted] node[left=0.1cm, yshift=0.1cm] {\shortstack[l]{$i_1$?$\phi_i^2$}} (l12)
    (l12) edge[dotted] node[left, yshift=0.35cm] {\shortstack[l]{$o_1$!$\phi_o^2$}} (l18)
    (l17) edge[dotted] node[above, yshift=0.35cm] {\shortstack[l]{$i_1$?$\phi_i^2$}} (l12)

    (l0) edge node[above, yshift=0.10cm] {\shortstack[l]{$i_1$?$\phi_i^3$}} (l5)
    (l5) edge node[above, yshift=0.10cm] {\shortstack[l]{$o_1$!$\phi_o^3$}} (l8)
    (l8) edge[dashdot] node[right, yshift=0.2cm] {\shortstack[l]{$i_2$?$\phi_i^4$}} (l10)
    (l10) edge[dashdot] node[right] {\shortstack[l]{$o_2$!$\phi_o^4$}} (l16)

    (l7) edge node[above, yshift=0.25cm] {\shortstack[l]{$i_1$?$\phi_i^3$}} (l14)
    (l14) edge node[right] {\shortstack[l]{$o_1$!$\phi_o^3$!}} (l20)
    (l17) edge node[above, yshift=1.3cm, xshift=5.2cm] {\shortstack[l]{$i_1$?$\phi_i^3$}} (l14)

    (l7) edge[dashdot] node[right=0.1, yshift=0.1cm] {\shortstack[l]{$i_2$?$\phi_i^4$}} (l13)
    (l13) edge[dashdot] node[right] {\shortstack[l]{$o_2$!$\phi_o^4$}} (l19)
    (l17) edge[dashdot] node[above, yshift=0.7cm, xshift=1.8cm] {\shortstack[l]{$i_2$?$\phi_i^4$}} (l13)
    (l18) edge[dashdot] node[right, yshift=-0.5cm, xshift=-0.5cm] {\shortstack[l]{$i_2$?$\phi_i^4$}} (l13)
    (l20) edge[dashdot] node[right, yshift=0.25cm] {\shortstack[l]{$i_2$?$\phi_i^4$}} (l13);

\end{tikzpicture}
}

\newcommand{\compositionCoverage}{
\begin{tikzpicture}[->, >=stealth, auto, node distance=3cm, thick,
    every state/.style={circle, draw, minimum size=1.2cm, thick, font=\huge},
    every node/.append style={font=\LARGE}, 
    initial text=, scale=0.52, transform shape]

\node[state, initial, initial where=above] (l0) {$l_0$};

\node[state, left=2.5cm of l0] (l3) {$l_3$};
\node[state, below=1.5cm of l0] (l4) {$l_4$};
\node[state, right=2.5cm of l0] (l5) {$l_5$};

\node[state, above=2.5cm of l5] (l1) {$l_1$};
\node[state, right=2.5cm of l1] (l2) {$l_2$};

\node[state, left=2.75cm of l3] (l6) {$l_6$};
\node[state, below=1.5cm of l4] (l7) {$l_7$};
\node[state, right=2.75cm of l5] (l8) {$l_8$};

\node[state, below left=of l6, red] (l9) {$l_9$};
\node[state, below right=of l8, red] (l10) {$l_{10}$};

\node[state, below left=2.7cm of l7, red] (l12) {$l_{12}$};
\node[state, left=4.5cm of l12, red] (l11) {$l_{11}$};
\node[state, below right=2.7cm of l7, red] (l13) {$l_{13}$};
\node[state, right=4.5cm of l13, red] (l14) {$l_{14}$};

\node[state, below=1.5cm of l9, red] (l15) {$l_{15}$};
\node[state, below=1.5cm of l10, red] (l16) {$l_{16}$};
\node[state, below=2.5cm of l11, red] (l17) {$l_{17}$};
\node[state, below=2.5cm of l12, red] (l18) {$l_{18}$};
\node[state, below=2.5cm of l13, red] (l19) {$l_{19}$};
\node[state, below=2.5cm of l14, red] (l20) {$l_{20}$};

\path
    (l0) edge[dashdot] node[above, yshift=0.2cm, xshift=-0.5cm] {\shortstack[l]{$i_2$?$\phi_i^4$}} (l1)
    (l1) edge[dashdot] node[above, yshift=0.10cm] {\shortstack[l]{$o_2$!$\phi_o^4$}} (l2)

    (l0) edge[dashed] node[right=0.2cm] {\shortstack[l]{$i_1$?$\phi_i^1$}} (l4)
    (l4) edge[dashed] node[right=0.2cm] {\shortstack[l]{$o_1$!$\phi_o^1$}} (l7)
    (l7) edge[dashed, red] node[above, yshift=0.25cm, red] {\shortstack[l]{$i_1$?$\phi_i^1$}} (l11)
    (l11) edge[dashed,red] node[right,red] {\shortstack[l]{$o_1$!$\phi_o^1$}} (l17)
    (l17) edge[dashed, bend left, red] node[left, red] {\shortstack[l]{$i_1$?$\phi_i^1$}} (l11)

    (l0) edge[dotted] node[above, yshift=0.10cm] {\shortstack[l]{$i_1$?$\phi_i^2$}} (l3)
    (l3) edge[dotted] node[above, yshift=0.10cm] {\shortstack[l]{$o_1$!$\phi_o^2$}} (l6)
    (l6) edge[dashdot, red] node[left, yshift=0.2cm, red] {\shortstack[l]{$i_2$?$\phi_i^4$}} (l9)
    (l9) edge[dashdot, red] node[left, red] {\shortstack[l]{$o_2$!$\phi_o^4$}} (l15)
    
    (l7) edge[dotted, red] node[left=0.1cm, yshift=0.1cm, red] {\shortstack[l]{$i_1$?$\phi_i^2$}} (l12)
    (l12) edge[dotted, red] node[left, yshift=0.35cm, red] {\shortstack[l]{$o_1$!$\phi_o^2$}} (l18)
    (l17) edge[dotted, red] node[above, yshift=0.35cm, red] {\shortstack[l]{$i_1$?$\phi_i^2$}} (l12)

    (l0) edge node[above, yshift=0.10cm] {\shortstack[l]{$i_1$?$\phi_i^3$}} (l5)
    (l5) edge node[above, yshift=0.10cm] {\shortstack[l]{$o_1$!$\phi_o^3$}} (l8)
    (l8) edge[dashdot, red] node[right, yshift=0.2cm, red] {\shortstack[l]{$i_2$?$\phi_i^4$}} (l10)
    (l10) edge[dashdot, red] node[right, red] {\shortstack[l]{$o_2$!$\phi_o^4$}} (l16)

    (l7) edge[red] node[above, yshift=0.25cm, red] {\shortstack[l]{$i_1$?$\phi_i^3$}} (l14)
    (l14) edge[red] node[right, red] {\shortstack[l]{$o_1$!$\phi_o^3$!}} (l20)
    (l17) edge[red] node[above, yshift=1.3cm, xshift=5.2cm, red] {\shortstack[l]{$i_1$?$\phi_i^3$}} (l14)

    (l7) edge[dashdot, red] node[right=0.1, yshift=0.1cm] {\shortstack[l]{$i_2$?$\phi_i^4$}} (l13)
    (l13) edge[dashdot, red] node[right] {\shortstack[l]{$o_2$!$\phi_o^4$}} (l19)
    (l17) edge[dashdot, red] node[above, yshift=0.7cm, xshift=1.8cm] {\shortstack[l]{$i_2$?$\phi_i^4$}} (l13)
    (l18) edge[dashdot, red] node[right, yshift=-0.5cm, xshift=-0.5cm] {\shortstack[l]{$i_2$?$\phi_i^4$}} (l13)
    (l20) edge[dashdot, red] node[right, yshift=0.25cm] {\shortstack[l]{$i_2$?$\phi_i^4$}} (l13);

\end{tikzpicture}
}

\runningauthor{Rodr\'iguez and Van den Bos}

\begin{abstract}

This paper combines methods from the fields of Model-Based Testing (MBT) and Behaviour-Driven Development (BDD) to define a testing approach with human-readable specifications and test cases, as in BDD, while using the modelling techniques and automatic test generation algorithms from MBT. We introduce PICKLES, a Precise Input and Control-flow Keyword-based Language for tEst Scenarios; an extension of Gherkin-style BDD scenarios, specified in structured natural language. We provide a bi-directional translation from Pickles scenarios to formal models, where both specifications and tests are human-readable, and a method to obtain a so-called master model combining all translated scenarios. Standard MBT algorithms can then be applied to automatically derive test cases from it. We implement a prototype of the translation and composition steps, which we use on an industrial case study: a software component from a traffic management system. With it, we illustrate the pipeline and show how Pickles can achieve significantly higher coverage with respect to BDD from the same set of scenarios.

\end{abstract}

\acknowledgment{The research is carried out as part of the NWO project \emph{Evidence-driven Black-box Checking (EVI)}, with project number OCENW.M.23.155. 
}

\begin{document}
\renewcommand{\sectionautorefname}{Section}
\renewcommand{\subsectionautorefname}{Section}
\renewcommand{\subsubsectionautorefname}{Section}

\maketitle
\urlstyle{rm}

\section{Introduction}
In critical systems, testing plays a central role in ensuring reliability and safety. Exhaustive testing is essential, as it directly strengthens confidence that the system behaves correctly under all expected conditions. Model-Based Testing (MBT) addresses this need by generating test cases from formal models of system behaviour, ensuring broad and systematic coverage. 


Despite its advantages, the adoption of MBT in industrial settings is limited. As pointed out by \cite{alegrothpractitioners2022}, practitioners recognize that effective MBT adoption depends on the active involvement of all stakeholders in modelling, including those without a technical background. A central barrier then is the difficulty of defining and maintaining formal system models, as expertise in formal specification techniques is required \cite{MBT-areas-tools-challenges}. Furthermore, communicating and discussing such models becomes challenging without a common language, understandable by all parties involved.


In practice, requirements are often specified in natural language \cite{francstatepractice2023}. This is the case for Behaviour-Driven Development (BDD), which has become popular in industry. In BDD, system behaviour is specified through executable examples, typically written as scenarios in a Given-When-Then format. These scenarios serve simultaneously as requirements documentation and as tests, and its human-readable nature facilitates the communication between experts from diverse domains. Nonetheless, example-based specifications can result in large test suites that are costly to maintain \cite{BDDAnalysis}. Moreover, defining requirements in natural language brings key challenges to overcome: ambiguity, inconsistency and incompleteness \cite{francstatepractice2023}.

\begin{figure}[b]
    \centering
    \includegraphics[width=1\linewidth]{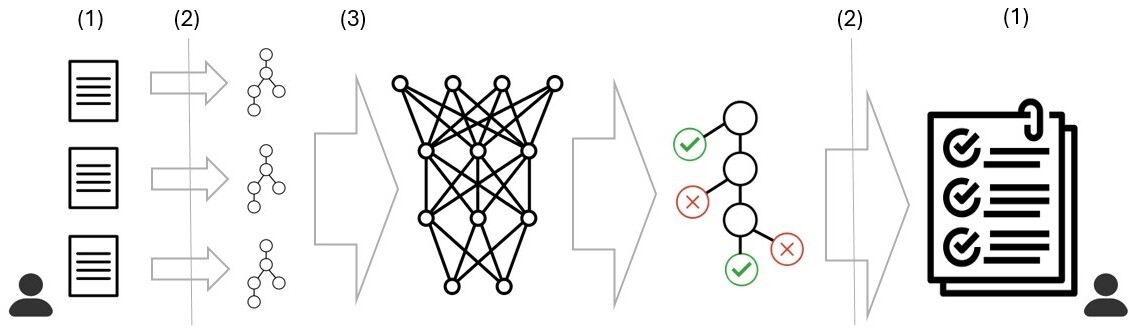}
    \caption{The Pickles testing pipeline: Pickles scenarios (1) are translated (2) into STS, which are then composed (3) into a master model. Formal test cases are derived from the master model, and translated back (2) to Pickles test cases (1).}
    \label{fig:pipeline}
\end{figure}

There is a clear tension in industrial practice: natural language is widely adopted due to its accessibility, however, it suffers from ambiguity and quality issues, while formal MBT models provide precision at the cost of readability and organizational acceptance. Existing approaches force practitioners to choose between these two opposites. To address this gap, we introduce PICKLES, a Precise Input and Control-flow Keyword-based Language for tEst Scenarios. In this framework, both the specifications and generated test cases remain human-readable in Pickles format, a Domain Specific Language grounded in BDD-style natural language constructs. A formal model is automatically derived from  Pickles specifications and composed into a master model; from it, test cases are generated by means of MBT algorithms, and then translated back to Pickles syntax. This way, we ensure both high-coverage, rigorous testing as well as maintainable, stakeholder-friendly artifacts. The contributions of this paper are as follows:




\begin{enumerate}[itemsep=1pt, topsep=2pt, parsep=0pt, partopsep=1pt]
    \item We introduce the PicklesDSL. This domain-specific language preserves the typical Given-When-Then structure used in BDD while introducing constructs for explicitly defining variable domains, constraints, and control flow. This way, Pickles specifications can capture a well-delineated set of requirements, preserving understandability for humans while ensuring unambiguous description of the desired behaviour. The syntax and semantics of this language are introduced in \autoref{sec:grammar}.
    \item We provide a bi-directional translation between PicklesDSL and formal models called Symbolic Transition Systems. Such models are derived \emph{automatically} from the scenarios, and vice versa; thus, test cases can also be human-readable, with the same terminology as the specifications. The translation from PicklesDSL specifications to STSs is introduced in \autoref{sec:grammar}, whereas \autoref{sec:test-cases} presents the translation from formal test cases back to PicklesDSL.
    \item We define how to compose a so-called \emph{master model} from a set of formal partial models, obtained via translation from a Pickles specification. The master model thus comprises a unified representation of the specified system behaviour, enabling the automatic derivation of tests that better reflect the system’s operational behaviour compared to isolated scenarios. This composition is introduced in \autoref{sec:modelcompos}.
    \item We implemented a prototype that automates the bi-directional translation, as well as the composition of the master model. This tool is introduced in \autoref{sec:implementation}.
    \item We demonstrate the applicability and advantages of our approach on a case study: a software component for traffic management, from the company Technolution. This case study is introduced in \autoref{sec:case-study} and serves as a running example throughout the paper to illustrate the introduced concepts and to discuss the benefits of the approach; this analysis is given in \autoref{sec:eval}.
    
\end{enumerate}
\autoref{fig:pipeline} depicts the complete Pickles testing pipeline, including where contributions (1), (2), and (3) take place.

\section{Preliminaries}

\subsection{Behaviour-Driven Development}\label{subsec:BDD}

Before introducing our case study, we briefly review Behaviour-Driven Development (BDD). BDD is a testing methodology centred around scenarios; concrete, example-based descriptions of how a system is expected to behave under specific conditions. Scenarios are written in Gherkin \cite{cucumber‐gherkin}, a domain-specific language that follows a Given–When–Then structure, representing a prescribed situation or context, execution steps of the system, and expected outcomes, respectively. A simple example Gherkin scenario for an ATM system is shown below:
\begin{verbatim}
Given the user accesses the ATM
And the user has a balance of 100 euros
When the user withdraws 20 euros
Then the user has a balance of 80 euros
\end{verbatim}
Scenarios can be transformed into automated test cases by developing concrete implementations for each sentence, often called \emph{step} (or otherwise  \emph{keyword}). Tools such as Cucumber \cite{cucumber}, Robot Framework \cite{robotframework}, and Behave \cite{behave} support this process by linking Gherkin specifications to executable test code.


\subsection{Variables, types and terms}\label{subsec:var-types}

Across the paper, we reason about variables; their types, domains, and functions over them. 
With $f : X \rightarrow Y$ we denote a total function from a domain $X$ to a codomain $Y$, with $Y ^ X$ being the set of functions from $X$ to $Y$.
With $T$, we denote the universe of variable \emph{types}, where $T_p = \{\texttt{boolean}, \texttt{integer}, \texttt{decimal}, \texttt{string}\}$ is the subset of \emph{primitive types}. We write $\Terms[\tau]{V}$ for a $\tau \in T$ to denote the set of $\tau$-typed terms over variables $V$, and $\mathcal{T}({V})$ to denote the set of terms over variables $V$ of any type. A term in $\Terms[\tau]{\emptyset}$ is called a \emph{ground term}. A (syntactical) ground term $t \in \Terms[\tau]{\emptyset}$ corresponds to a (semantical) value $u \in \valueuniv_\tau$, where $\valueuniv_\tau$ denotes the universe of values of type $\tau$. With $\valueuniv$ we denote the set of all possible values, of any type.

We also define some auxiliary functions. Function \( \mathsf{type}_v : V \to \tau\) returns the \emph{type} (data type) of a variable $v \in V$; \( \mathsf{type}_u : \valueuniv \to \tau\) does the same for a value $u \in \valueuniv$. The function \(\mathsf{dom} : V \to \mathcal{P}(\valueuniv_\tau)\) yields the \emph{domain} of a given variable, i.e. the set of values it can take; for any $v \in V$, \(\domfun{v} \subseteq \valueuniv_ {\sortfun{v}}\). 

\subsection{Symbolic Transition Systems}\label{subsec:sts}

We will express the semantics of our DSL in terms of Symbolic Transition Systems (STS), as defined in \cite{covbasedtesting}. This formalism extends Labelled Transition Systems (LTS) with data constructs. In particular, the transition of an STS, extended with data constructs, is called a \emph{switch}. An STS has global variables, called \emph{location variables}, as well as variables local to switches, called \emph{parameters}.  Switches consist of an input or output \emph{gate}, and may be augmented with \emph{guards} over parameters and location variables, as well as with \emph{assignments}, that is, updates to the values of location variables. 

\begin{definition}\label{def:sts}
A \emph{Symbolic Transition System (STS)}, denoted by $\STS$, is a tuple $(\locset, l_{0}, \vlset, \vpset, \Gamma,\interact, \switchrel)$ where:

\begin{itemize}[label=--]
    \item $\locset$ is a finite set of locations,
    \item $l_{0} \in \locset$ is the initial location,
    \item $\vlset$ is a finite set of location variables,
    \item $\vpset$ is a finite set of parameters,
    \item $\Gamma = \Gamma_{I} \cup \Gamma_{O}$ is a finite set of input and output gates, with $\Gamma_{I} \cap \Gamma_{O} = \emptyset$,
    \item The interaction function $\interact: \Gamma \rightarrow \vpset^*$ maps each $\gamma \in \Gamma$, to a sequence $\bar{p} \in \vpset^*$ of distinct parameters. Often we use set $\interact \subseteq \Gamma \times \vpset^*$ instead of function $\interact: \Gamma \rightarrow \vpset^*$, and refer to interaction $(\gamma,\bar{p}) \in \interact$ instead of writing $\interact(\gamma)=\bar{p}$,  
    \item $\switchrel \subseteq \locset \times \interact \times \mathcal{T}_{Bool}(\vlset \cup \vpset) \times \mathcal{T}(\vlset \cup \vpset)^{\vlset} \times \locset$ is the switch relation.
\end{itemize}
\end{definition}
We refer to the elements of a switch $(l,\alpha,\phi,\psi,l') \in \switchrel$, with source location, interaction, guard, assignment, and target location, respectively, and to the elements of interaction $\alpha = (\gamma,p_0\dots p_k)$, as gate and parameters. We require for each switch that $\phi \in \boolterm(\vlset \cup \{p_0,\dots, p_k\})$, and $\psi \in \term(\vlset\cup \{p_0,\dots, p_k\})^{\vlset}$, i.e. guards and assignments can use any local variable, and only parameters of their switch.

We distinguish input and output switches by their gate, as $\switchrel_I = \{ r \in \switchrel \;|\; \gamma \in \Gamma_I \}$ and $\switchrel_O = \{ r \in \switchrel \;|\; \gamma \in \Gamma_O \}$.
Moreover, we use superscripts to refer to an element of a given STS, e.g., $\locset^i$ refers to the set of location variables of $\STS_i$, and $l^i_0$ to its initial location. In addition, we  write $l\xrightarrow{\alpha,\phi,\psi}l'$ to express that $(l,\alpha,\phi,\psi,l') \in \switchrel$, i.e. it is a defined switch, and we write $l\not\xrightarrow{\alpha,\phi,\psi}$ to denote that for any $l'\in \locset$ we have $(l,\alpha,\phi,\psi,l') \notin \switchrel$, i.e. the switch does not exist.
Note that we do not include a function that assigns initial values to the location variables, as usual for STS \cite{covbasedtesting}; we define such values when defining test cases in \autoref{sec:formaltest}.

\subsection{Context-free Grammar} \label{sec:ebnf}

%
%
In this work, we describe the Pickles DSL via a context-free grammar $P$. To describe it, we adopt a regex-like variant of Extended Backus–Naur Form (EBNF). In it, non-terminals are enclosed in angle brackets or written in \textsf{sans-serif font}, while terminals have \texttt{typewriter font}. We also use the standard symbols ``$*$'', ``$+$'', ``$?$'', ``$|$'', and parentheses in production rules.

Throughout the paper, we denote by $\nonterm{s}_P$ the \emph{sub-language} of $P$ generated by a grammar symbol $\nonterm{s}$, that is, the set of all words derivable from $\nonterm{s}$. In addition, we use the notation $\semmap[s]{\word{s}}$ to denote the \emph{semantic interpretation} of $\word{s}$, where $\word{s}$ is a word in $\nonterm{s}_P$. As we define our grammar's semantics by mapping its elements to those of an STS, we may also refer to $\semmap[s]{\word{s}}$ as the \emph{semantic mapping} of the word $\word{s}$.

\section{Introduction to the Case Study}\label{sec:case-study}

To test the applicability and benefits of our approach, we apply it to a software component of the Dutch company Technolution B.V. 
This component is part of a traffic management system; it receives telemetry values from sensors distributed across a road section. These sensors aim to detect abnormal events, such as drivers driving at a low speed or standing still. Based on the events' location, especially with regards to a critical subsection of the road, status information is sent to the user interface.

In \autoref{lst:casestudy} we present a specification suite for the system, expressed in PicklesDSL. The suite consists of four scenarios and their domain, expressed in the variable settings prelude. These scenarios were manually derived from a set of BDD test cases, implemented in Cucumber, provided by Technolution. The resulting Pickles scenarios do not correspond one-to-one with the original test cases; instead, during their definition, we incorporated additional domain-knowledge information to generalize the expected behaviour.

\begin{lstlisting}[language=PicklesDSL, caption={Pickles Specification Suite for the Technolution case study.},label=lst:casestudy, numbers=right][t]
Variable Settings
"availability" is a string with range (@\textcolor{black}{\{AV, PART AV, NOT AV\}}@)
"enabledness" is a boolean with range (@\textcolor{black}{\{true, false\}}@)
"faulty detectors" is an array of at most (@\textcolor{black}{3} \label{line:fd-def-start}@)elements where each element is a structure with attributes "lane", "lenght position" such that:
  "lane" is an integer with range (@\textcolor{black}{[1,3]}@)
  "length position" is a decimal with range (@\textcolor{black}{(1.0,3.0)}\label{line:fd-def-end}@)
"critical section lane" is an integer with range (@\textcolor{black}{[1,3]}@)
"critical section start" is a decimal with range (@\textcolor{black}{(1.0,3.0)}@)
"critical section end" is a decimal with range (@\textcolor{black}{(1.0,3.0)}@)

Scenario (@\textcolor{black}{01: faulty detectors outside the critical section}@)
(@\textcolor{dkblue}{Given the system is in its initial state} \label{line:given-ini}@)
And the (@\textcolor{black}{user interface displays information}@) (@\textcolor{black}{on}@) "enabledness", "availability" such that:
  "enabledness" is equal to (@\textcolor{black}{true}@) AND
  "availability" is equal to (@\textcolor{black}{AV} \label{line:given-end}@)
When (@\textcolor{black}{the controller detects}@) "faulty detectors" such that:
  "faulty detectors" has all elements where each element has attributes such that:
    "lane" is not equal to "critical section lane" OR
    "lenght position" is lower or equal than "critical section start" OR
    "lenght position" is greater or equal than "critical section end"
Then (@\textcolor{black}{the user interface displays}@) "availability" equal to (@\textcolor{black}{AV} \label{line:synsugar}@)

Scenario (@\textcolor{black}{02: one faulty detector in the critical section}@)
(@\textcolor{dkblue}{Given the system is in its initial state}@)
And the (@\textcolor{black}{user interface displays information}@) (@\textcolor{black}{on}@) "enabledness", "availability" such that:
  "enabledness" is equal to (@\textcolor{black}{true}@) AND
  "availability" is equal to (@\textcolor{black}{AV}@)
When (@\textcolor{black}{the controller detects} \label{line:when-s2-start}@) "faulty detectors" such that:
  "faulty detectors" (@\textcolor{black}{has}@) exactly (@\textcolor{black}{1}\label{line:cond-when-s2-start}@) elements where each element has attributes such that:
    "lane" is equal to "critical section lane" AND
    "lenght position" is greater than "critical section start" AND
    "lenght position" is lower than "critical section end"(@\label{line:when-s2-end}@)
Then (@\textcolor{black}{the user interface displays}@) "availability" equal to (@\textcolor{black}{PART AV} \label{line:myline}@)

Scenario (@\textcolor{black}{03: at least two faulty detectors in the critical section}@)
(@\textcolor{dkblue}{Given the system is in its initial state}@)
And the (@\textcolor{black}{user interface displays information}@) (@\textcolor{black}{on}@) "enabledness", "availability" such that:
  "enabledness" is equal to (@\textcolor{black}{true}@) AND
  "availability" is equal to (@\textcolor{black}{AV}@)
When (@\textcolor{black}{the controller detects}@) "faulty detectors" such that:
  "faulty detectors" (@\textcolor{black}{has}@) at least (@\textcolor{black}{2}@) elements where each element has attributes such that:
    "lane" is equal to "critical section lane" AND
    "lenght position" is greater than "critical section start" AND
    "lenght position" is lower than "critical section end"
Then (@\textcolor{black}{the user interface displays}@) "availability" equal to (@\textcolor{black}{NOT AV}\label{line:S3then}@)

Scenario (@\textcolor{black}{04: lost controller access disables the system}@)
(@\textcolor{dkblue}{Given the system is in its initial state}@)
And (@\textcolor{black}{the user interface displays information on}@) "enabledness" such that:
  "enabledness" is equal to (@\textcolor{black}{true}@)
When (@\textcolor{black}{the controller access is lost}@)
Then (@\textcolor{black}{the user interface reports status}@) "enabledness" equal to (@\textcolor{black}{false}@)
\end{lstlisting}

These scenarios define the expected system behaviour when malfunctioning detectors are reported, particularly those located within a critical road section. Each sensor and critical section is localized by means of a discrete lane and a longitudinal position, where the latter represents a continuous value orthogonal to the lanes. At most 3 sensors may fail simultaneously. If there are any number of failing detectors but none of them is in the critical section, the system is available (Scenario 1). If at least one failing sensor is within the critical section, that is, in the same lane and length position, the system reports it is partially available (Scenario 2). If two or more failing detectors match the critical section position, it must report the unavailability of the system (Scenario 3). Finally, if the controller (i.e. whole system) is not accessible at some point, the system is disabled and no other action can be performed afterwards (Scenario 4).

\section{Pickles grammar and translation to STS}\label{sec:grammar}

In this section, we introduce the syntax of the Pickles DSL by means of defining a context-free grammar. We also define its semantics by mapping each of the grammar's elements to elements of a Symbolic Transition System. Throughout this section, we refer to the Pickles scenarios presented in \autoref{sec:case-study} to illustrate the introduced concepts.

\subsection{Assumptions}

While interpretations of the Given–When–Then steps in Gherkin scenarios (see \autoref{subsec:BDD}) vary across authors, we constrain ours as follows. The \textbf{Given} clause represents the precondition of the scenario, i.e. the state of the System Under Test (SUT) required to execute the scenario. The \textbf{When} clause denotes an interaction with the SUT initiated by an external actor, for example, a user. Internal transitions and actions executed by the SUT are excluded. Finally, the \textbf{Then} clause represents an observable action performed by the SUT. We assume that all scenarios have at least one When and one Then step.

\subsection{Syntax and Semantics}

This section presents the concrete mapping from the PicklesDSL grammar, $P$, to an STS. 

As shown in \autoref{lst:casestudy}, an instance of the PicklesDSL grammar, called \emph{specification suite}, consists of a set of Scenarios and a Variable Settings block. Each scenario describes a functionality in the typical Given–When–Then structure used in BDD; each scenario maps to a STS. As part of this process, each clause preceded by the \textcolor{dkblue}{\texttt{Given}} token is interpreted as a precondition of the scenario; in the STS, this becomes an additional guard over the first switch. Switches are further defined by \textcolor{dkblue}{\texttt{When}} and \textcolor{dkblue}{\texttt{Then}} clauses, which are mapped to input and output switches, respectively. Scenarios may be parametrized using \textcolor{dkgreen}{\texttt{variables}} and guarded by conditions over them. All variables used across scenarios must be declared in the Variable Settings prelude, where their types (e.g. \textcolor{goldenrod}{\texttt{integer}}) and ranges (e.g. \texttt{[1, 3]}) are defined. Variables may appear in any step. Conditions over variables can be specified to constrain their admissible values through the use of operators (e.g., \textcolor{dkred}{\texttt{equal}}, \textcolor{dkred}{\texttt{between}}, \textcolor{dkred}{\texttt{lower than}}). Multiple conditions can be connected through logical operators such as \textcolor{purple}{\texttt{AND}} or \textcolor{purple}{\texttt{OR}}. These conditions are mapped to switch guards.



The subsequent subsections define how elements of the PicklesDSL grammar are mapped to STS elements. We first cover the high-level elements before addressing the specific syntax and semantics of the Variable Settings Block, Guard Blocks, and the Given, When, and Then blocks. The corresponding rules for the inverse process, translating formal test cases back into PicklesDSL syntax, will be introduced in \autoref{sec:formaltest}.

\subsubsection{Specification Suite}\label{subsubsec:SpecSuite}

The non-terminal $\nonterm{SpecSuite}$ is the start symbol of our grammar. Its production rule states that a $\nonterm{VarSetBlock}$ is followed by one or more \textsf{Scenarios}. Each scenario consists of a textual description (only for readability as it is not taken into account for the mapping to STS), and of $\nonterm{Given}$, $\nonterm{When}$, and $\nonterm{Then}$ words, capturing initial conditions, inputs, and expected outputs. These may reference variables, and include guards to specify conditions. 

\begin{grammar}
\caption{Pickles Grammar $P$: Specification Suite}\label{grammar:overview}
\setlength{\jot}{1pt} 
\vspace{-7pt}
\begin{smallmath}
\langle \text{SpecSuite} \rangle &::= \langle \text{VarSetBlock} \rangle ~ \langle \text{Scenario} \rangle+\\
\langle \text{VarSetBlock} \rangle &::= \text{\texttt{Variable Settings}}~ (\text{\texttt{"}}\langle \text{VarID} \rangle\text{\texttt{"}}~ \\
&\phantom{:::=} ~\langle \text{TypeDesc} \rangle)\text{*}\\
\langle \text{Scenario} \rangle &::= \text{\texttt{Scenario}}~ \langle \text{Str} \rangle ~\langle \text{Given} \rangle?~ \langle \text{When} \rangle ~\langle \text{Then} \rangle\\
\langle \text{Given} \rangle &::= \text{\texttt{Given}}~\langle \text{InitialCond} \rangle?~ \langle \text{Str} \rangle?~\text{\texttt{such}}\\
&\phantom{:::=}~ \text{\texttt{that:}}~\langle \text{GuardBlock} \rangle\\
\langle \text{When} \rangle &::= \text{\texttt{When}}~ \langle \text{InStep} \rangle ~(\text{\texttt{And}}~ \langle \text{InStep} \rangle)\text{*} \\
\langle \text{InStep} \rangle &::= \langle \text{InAction} \rangle ~((\text{\texttt{"}}\langle \text{VarID} \rangle\text{\texttt{"}})\text{+}~ \text{\texttt{such}}\\
&\phantom{:::=}~ \text{\texttt{that:}}~ \langle \text{GuardBlock} \rangle)?\\
\langle \text{Then} \rangle &::= \text{\texttt{Then}}~ \langle \text{OutStep} \rangle~ (\text{\texttt{And}}~ \langle \text{OutStep} \rangle)\text{*}\\
\langle \text{OutStep} \rangle &::= \langle \text{OutAction} \rangle ~((\text{\texttt{"}}\langle \text{VarID} \rangle\text{\texttt{"}})\text{+}~ \text{\texttt{such}}\\ 
&\phantom{:::=}~ \text{\texttt{that:}}~ \langle \text{GuardBlock} \rangle)?
\end{smallmath}
\vspace{-7pt}
\end{grammar}

Given $n$ words $\word{s} \in \nonterm{Scenario}_P$ in the $\nonterm{SpecSuite}$, we define a set of $n$ STSs $\STSset = \{\STS_1, ..., \STS_n\}$. Within $\STSset$, we distinguish two subsets: primary (or initial) STSs, denoted by $\STSset_{p}$, which correspond to scenarios where the token $\nonterm{InitialCond}$ is present; and secondary STSs, denoted by $\STSset_{s}$, which correspond to scenarios where it is absent. Primary STSs can be executed immediately after system initialization, whereas secondary STSs can only be executed following another STS, provided that their guard condition is satisfied. The resulting set is then composed into a single master model, $\STS_{master}$; this construction is detailed in \autoref{sec:modelcompos}. 

\subsubsection{Variable Settings Block}\label{subsubsec:VarSetBlock}

The $\nonterm{VarSetBlock}$ introduces the data types and admissible ranges for all variables in the specification suite. Variables are classified by their data types in primitives (see \autoref{subsec:var-types}), arrays, or structures. 
Grammar~\autoref{grammar:VarSetBlock} shows the production rules for the $\nonterm{VarSetBlock}$ non-terminal.

\begin{grammar}[h]
\caption{Pickles Grammar $P$: Variable Settings Block}\label{grammar:VarSetBlock}
\setlength{\jot}{1pt}
\vspace{-7pt}
\begin{smallmath}
\langle \text{VarSetBlock} \rangle &::= \text{\texttt{Variable Settings}}~ (\text{\texttt{"}}\langle \text{VarID} \rangle\text{\texttt{"}}~ \\
&\phantom{:::=} ~\langle \text{TypeDesc} \rangle)\text{*}\\
\langle \text{TypeDesc} \rangle &::= \langle \text{Primitive} \rangle ~|~ \langle \text{Array} \rangle ~|~ \langle \text{Struct} \rangle\\
\langle \text{Primitive} \rangle &::= \text{\texttt{is a }} \langle \text{PrimType} \rangle ~\text{\texttt{with range}}~ \langle \text{Range} \rangle\\
\langle \text{PrimType} \rangle &::= \text{\texttt{boolean}}~ \mid ~\text{\texttt{string}}~ \mid  ~\text{\texttt{integer}}~ \mid ~\text{\texttt{decimal}}\\
\langle \text{Array} \rangle &::= \text{\texttt{is an array of}}~ \langle \text{Cardinality} \rangle ~\text{\texttt{elements }}\\
&\phantom{:::=} ~\text{\texttt{where each element}}~ \langle \text{TypeDesc} \rangle\\
\langle \text{Cardinality} \rangle &::= \text{\texttt{at most}}~ \langle \text{N} \rangle ~|~ \text{\texttt{exactly}}~ \langle \text{N} \rangle ~|~\\
&\phantom{:::=} ~ \text{\texttt{between}}~ \langle \text{N} \rangle ~\text{\texttt{and}}~ \langle \text{N} \rangle \\
\langle \text{Struct} \rangle &::= \text{\texttt{is a structure with attributes}}~ \\
&\phantom{:::=}~\text{\texttt{"}}\langle \text{AttrID} \rangle\text{\texttt{"}}~(\text{\texttt{,}}~ \text{\texttt{"}}\langle \text{AttrID} \rangle\text{\texttt{"}})\text{*}~
\text{\texttt{such that:}}\\
&\phantom{:::=}~\langle \text{AttrDesc} \rangle\text{+}\\
\langle \text{AttrDesc} \rangle &::= \text{\texttt{"}}\langle \text{AttrID} \rangle\text{\texttt{"}}~ \langle \text{TypeDesc} \rangle\\
\end{smallmath}
\vspace{-7pt}
\end{grammar}

To describe the semantics of the $\nonterm{VarSetBlock}$ non-terminal, we first define those of $\nonterm{Type Desc}$ and $\nonterm{VarID}$. 

\paragraph{Type Description}

Words derived from the $\nonterm{TypeDesc}$ non-terminal define both the data type and the range for any given variable, which will be mapped to sorts and domains of STS variables. The most fundamental construction is the primitive type, which serves as the building block for more complex definitions. Arrays and structures are constructed through a recursive process, defining the types of their elements or attributes until they eventually resolve into primitive data types. For instance, a structure may be composed of multiple attributes, such as a combination of an integer and an array of strings.

Let $\semmap[TypeDesc]{\cdot}: \nonterm{TypeDesc}_P \to V$ map words in the sub-language $\nonterm{TypeDesc}_P$ to sorted variables. Next, we introduce its definition for primitives, arrays and structures, respectively.

\paragraph{Primitive Definitions}
For word $w_{\text{p}} \in \nonterm{Primitive}_P$ of the form:
\begin{smallmath}
w_{\text{p}} = \text{\texttt{is a $t$ with range}}\ r 
\end{smallmath}
mapping $\semmap[Primitive]{w_{\text{p}}}$ yields a variable $v$, with type $\semmap[PrimType]{t}$, and domain $\semmap[Range]{r}$. 
Here, $\semmap[PrimType]{\cdot} : \nonterm{PrimType}_P \to T_p$ maps each $\nonterm{PrimType}$ word to the primitive data type it represents, and $\semmap[Range]{\cdot} : \nonterm{Range}_P \to \mathcal{P}(\valueuniv_\tau), \tau \in T_p$ maps range words to sets of primitive values. 
We do not further define these two functions, and assume that an appropriate representation can be defined for STS, and simply require that \(\mathrm{dom}(v) \subseteq \valueuniv_{\mathrm{type}(v)}\). For example, the word \texttt{is an integer with range \{a,b,c\}} is not valid as the type and range of the variable are incompatible.

\paragraph{Array Definitions}
 An \texttt{array} is a data type that represents a collection of elements of a common type. 

Consider a word $w_{\text{a}} \in \nonterm{Array}_P$ of the form:
\begin{smallmath}
    w_{\text{a}} &= \text{\texttt{is an array of $w_{\text{c}}$ elements such that}}\\
    &\phantom{=}~\text{\texttt{each element }} w_{\text{e}}
\end{smallmath}
This word is mapped to an \texttt{array} where all elements are of the type given by the word $w_{\text{e}} \in \nonterm{TypeDesc}_P$. Its domain is the set of all possible arrays of given cardinality $w_{\text{c}}$, that can be formed with elements in the domain of $w_{\text{e}}$. 
For example, the word \texttt{is an array of exactly 2 elements such that each element is a string with range \{a,b,c\}} yields a variable $v$, which is an array of strings, with as domain the set of all arrays of length 2 consisting of strings \text{\texttt{a}}, \text{\texttt{b}} and \texttt{c}. 



\paragraph{Structure Definitions}
The \texttt{structure} data type denotes a mapping from unique keys to sorted values. 
We introduce $\semmap[AttrID]{\cdot}$, a function for mapping words from the $\nonterm{AttrID}_P$ sub-language to keys of a structure. There is no constructive definition for this function; instead, we assume that each occurrence of a $\nonterm{AttrID}$ consistently refers to the same key, and that this correspondence is recorded for later reuse.

Consider a word $w_{\text{s}} \in \nonterm{Struct}_P$ of the form:
\begin{smallmath}
\word{s} &= \text{\texttt{is a structure with attributes }} \\
&\phantom{=}~\text{\texttt{"}}\word[0]{att}\text{\texttt{"}}, \dots, \text{\texttt{"}}\word[n]{att}\text{\texttt{"}} \text{\texttt{ such that: }}\\
&\phantom{=}~\text{\texttt{"}}\word[0]{att}\text{\texttt{"}} ~\word[0]{td}, \; \dots,\; \text{\texttt{"}}\word[n]{att}\text{\texttt{"}} ~\word[n]{td}
\end{smallmath}
with $\word[0]{att},\dots,\word[n]{att} \in \nonterm{AttrID}_P$, and $\word[0]{td},\dots,\word[n]{td} \in \nonterm{TypeDesc}_P$. Then $\semmap[Struct]{w_{\text{s}}}$ yields a \texttt{structure}, where the i-th attribute key is given by $\word[i]{att}$ and the i-th attribute's type description is given by $\word[i]{td}$, denoting both its sort and range. For example, the word \texttt{is a structure with attributes "AttrA", "AttrB" such that: "AttrA" is an integer with range [1], "AttrB" is a string with range \{a,b\}} yields a variable of type \texttt{structure}, and its domain is the set: $\{\{\text{\texttt{"AttrA"}}: 1, \text{\texttt{"AttrB"}}: \text{\texttt{a}}\}, \{\text{\texttt{"AttrA"}}: 1, \text{\texttt{"AttrB"}}: \text{\texttt{b}}\}\}$.



\paragraph{Variable ID}
In the $\nonterm{VarSetBlock}$, words derived from the $\nonterm{VarID}$ non-terminal denote both \emph{location variables} and \emph{gate parameters} in the resulting STS. They are not distinguished in the $\nonterm{VarSetBlock}$; instead, the mapping resolves this when constructing the STS by defining simultaneously a location variable and a gate parameter from the same identifier, both with the same type and domain. 

The function $\semmap[VarID]\cdot : \nonterm{VarID}_P \rightarrow \vlset \times \vpset$ maps words in $\nonterm{VarID}$ to tuples of location variable, gate parameter. We also define the projection functions $\pi_{l} : \vlset \times \vpset \rightarrow \vlset$ and $\pi_{p} : \vpset \times \vlset \rightarrow \vpset$. We assume the correspondence between identifications and variables is always consistent and recorded.

\paragraph{Variable Settings Block}
In the $\nonterm{VarSetBlock}$ restrictions for the datatypes and ranges of all variables are introduced. As location variables and gate parameters with the same identifier must be consistent, at all times for an identifier $id$ then $\sortfun{\pi_{l}(\semmap[VarID]{id})} = \sortfun{\pi_{p}(\semmap[VarID]{id})}$ holds.
Then, for some word $w_{\text{vs}} \in \nonterm{VarSetBlock}_P$ of the form:
\begin{smallmath}
    \word{vs} = \text{\texttt{Variable Settings }} \text{\texttt{"}} w_{\text{id}_0} \text{\texttt{" }} w_{\text{td}_0}\; \dots \text{\texttt{ "}} w_{\text{id}_m} \text{\texttt{" }} w_{\text{td}_m}
\end{smallmath}

with $w_{\text{td}_0}, \dots, w_{\text{td}_m} \in \textsf{TypeDescr}_P$ and $w_{\text{id}_0}, \dots, w_{\text{id}_m} \in \textsf{VarID}_P$, we define:

\begin{smallmath}
    \vlset &= \{ \pi_{l}(\semmap[VarID]{w_{\text{id}_j}}) \;|\; 0 \leq j \leq m\}\\
    \vpset &= \{ \pi_{p}(\semmap[VarID]{w_{\text{id}_j}}) \;|\; 0 \leq j \leq m\}
\end{smallmath}
as the sets of location variables and gate parameters shared by all the STSs derived from the corresponding specification suite. For all $v_j \in \vlset$, $p_j \in \vpset$:
\begin{smallmath}
    \sortfun{v_j} &= \sortfun{p_j} = \sortfun{\semmap[X_j]{w_{\text{td}_j}}} \\\ \domfun{v_j} &= \domfun{p_j} = \domfun{\semmap[X_j]{w_{\text{td}_j}}}
\end{smallmath}
where $\textsf{X}_j$ denotes the syntactic category of $w_{\text{td}_j}$: \textsf{Primitive}, \textsf{Array} or \textsf{Struct}, as defined in Grammar~\autoref{grammar:VarSetBlock}.

All words derived from $\nonterm{VarID}$ used across the Specification Suite must be introduced in $\nonterm{VarSetBlock}$. We also restrict $\nonterm{VarID}_P \cap \nonterm{AttrID}_P = \emptyset$, meaning that variables and structures' attributes may not have identical identifiers.

\begin{example}
    We consider the example introduced in \autoref{lst:casestudy}, in particular, the definition of the variable \emph{faulty detectors} (see lines \ref{line:fd-def-start}-\ref{line:fd-def-end}).
    \autoref{tab:variables} presents a summary of all the variable naming that we will conserve throughout the paper. In addition, we consider $\semmap[AttrID]{\text{lane}} = l$ and $
        \semmap[AttrID]{\text{length position}} = \mathit{lp}$
     The mapping of \emph{faulty detectors} yields both a location variable, $v_{\text{fd}}$, and a gate parameter $p_{\text{fd}}$. In both cases, it is an array of one to three structures, each with two attributes: $l$, an integer in $\{1,2,3\}$, and $\mathit{lp}$, a decimal strictly between $1.0$ and $3.0$. 
    
\vspace{-8pt}
\begin{table}[h]
\centering
\small
\caption{Variables in \autoref{lst:casestudy} and their identifiers.}\label{tab:variables}
\renewcommand{\arraystretch}{1.25}
\begin{tabular}{ 
>{\centering\arraybackslash}m{3.5cm}  
                 >{\centering\arraybackslash}m{2cm}  
                 >{\centering\arraybackslash}m{2cm}}  
\hline
id & $\pi_{l}(\semmap[VarID]{id})$ & $\pi_{p}(\semmap[VarID]{id})$ \\ 
\hline
availability              & $v_{\text{av}}$   & $p_{\text{av}}$ \\
enabledness                   & $v_{\text{en}}$   & $p_{\text{en}}$\\
faulty detectors & $v_{\text{fd}}$ & $p_{\text{fd}}$ \\
critical section lane     & $v_{\text{cl}}$   & $p_{\text{cl}}$\\
critical section start    & $v_{\text{cs}}$   & $p_{\text{cs}}$\\
critical section end      & $v_{\text{ce}}$   & $p_{\text{ce}}$\\
\hline
\end{tabular}
\end{table}
\vspace{-8pt}
\end{example}

\subsubsection{Guard Block}\label{subsubsec:GuardBlock}

As mentioned previously, the Pickles specification suite does not define system behaviour as a mere sequence of steps. These steps can be augmented with variables and guarded conditions, enabling parametrization. This shifts the focus from isolated examples to a generalized behavioural description of the system. Input values can be chosen for variables using some test selection strategy, e.g. equivalence partitioning.
The same principle applies to outputs: not only can we define a single exact expected value, but we can also specify predicates over variables, to e.g. specify acceptable ranges. 

A $\nonterm{GuardBlock}$ word defines restrictions for the values of scenarios' variables in each step. A guard is local to a switch; however, it must be compatible with the variable's type and domain, as defined in the variable settings prelude. The production rules for $\nonterm{GuardBlock}$ are presented in Grammar \autoref{grammar:guard-block}.

\begin{grammar}[b]
\caption{Pickles Grammar $P$: Guard Block}\label{grammar:guard-block}
\setlength{\jot}{1pt}
\vspace{-7pt}
\begin{smallmath}
\langle \text{GuardBlock} \rangle &::=  \text{\texttt{"}}\langle \text{VarID} \rangle \text{\texttt{"}} \langle \text{Guard} \rangle (\langle \text{ConjOp} \rangle \text{\texttt{"}}\langle \text{VarID} \rangle \text{\texttt{"}}\\
&\phantom{:::=}~\langle \text{Guard} \rangle)\text{*}\\
\langle \text{Guard} \rangle &::= \langle \text{PrimGuard} \rangle ~|~ \langle \text{ArrayGuard} \rangle ~|~ \langle \text{StructGuard} \rangle\\
\langle \text{PrimGuard} \rangle &::= \text{\texttt{is}}~ \langle \text{Op} \rangle ~(\langle \text{ExpValue} \rangle ~|~ \langle \text{VarRef} \rangle)\\
\langle \text{Op} \rangle &::= \text{\texttt{equal to}} ~|~ \text{\texttt{greater than}} ~|~ \text{\texttt{lower than}} ~|~ \\ &\text{\texttt{lower or equal than}} ~|~ \text{\texttt{greater or equal than}}\\
\langle \text{ArrayGuard} \rangle &::= \text{\texttt{has}}~ \langle \text{Quantifier} \rangle \langle \text{N} \rangle ~\text{\texttt{elements where}}\\ 
&\phantom{:::=} ~\text{\texttt{each element}}~ \langle \text{Guard} \rangle\\
\langle \text{Quantifier} \rangle &::= \text{\texttt{at least}} ~|~ \text{\texttt{at most}} ~|~ \text{\texttt{exactly}} ~|~ \text{\texttt{all}}\\
\langle \text{StructGuard} \rangle &::= \text{\texttt{has attributes such that:}}~ \langle \text{AttrGuard} \rangle\\
&\phantom{:::=}~(\langle \text{ConjOp} \rangle \langle \text{AttrGuard} \rangle)\text{*}\\
\langle \text{AttrGuard} \rangle &::= \text{\texttt{"}}\langle \text{AttrID} \rangle \text{\texttt{"}}~\langle \text{Guard} \rangle\\
\langle \text{VarRef} \rangle &::= \text{\texttt{stored}}?~ \text{\texttt{"}}\langle \text{VarID} \rangle\text{\texttt{"}}\\
\langle \text{ConjOp} \rangle &::= \text{\texttt{AND}} ~|~ \text{\texttt{OR}}\\
\langle \text{ExpValue} \rangle &::= \langle \text{Range} \rangle ~|~ \langle \text{N} \rangle ~|~ \langle \text{R} \rangle ~|~ \langle \text{Str} \rangle ~|~ \langle \text{Aexp} \rangle
\end{smallmath}
\vspace{-7pt}
\end{grammar}

We can now define $\semmap[Guard Block]{\cdot} : \nonterm{Guard Block}_P \to \boolterm{(\vpset \cup \vlset)}$, a function that maps words in $\nonterm{Guard Block}_P$ to predicates over location variables and gate parameters. 

Consider a word $w_{\text{gb}} \in \nonterm{Guard Block}_P$:
\vspace{-1ex}
\begin{smallmath}
    \word{gb} = \text{\texttt{"}}\word[0]{id}\text{\texttt{" }} \word[0]{g} \; \word[0]{c} \; \dots \; \word[n-1]{c} \text{\texttt{ "}}\word[n]{id}\text{\texttt{" }} \word[n]{g}
\end{smallmath}
such that $\word[0]{id} \dots \word[n]{id} \in \nonterm{VarID}_P$, $\word[0]{c} \dots \word[n-1]{c} \in \nonterm{ConjOp}_P$ and $\word[0]{g} \dots \word[n-1]{g} \in \nonterm{Guard}_P$. $\semmap[Guard Block]{\word{gb}}$ yields a predicate over the variables represented by $\word[0]{id} \dots \word[n]{id}$. Such predicate is built by connecting sub-formulas (each of them potentially over different variables) by means of logic operators, denoted by words $\word[0]{c} \dots \word[n-1]{c}$. Here, $\semmap[Conjuntion]{\cdot} : \nonterm{ConjOp}_P \to \{\land, \lor\}$ maps \texttt{AND} and \texttt{OR} to $\land$ and $\lor$ respectively.

We distinguish two cases in the semantics of guards. If a word in $\nonterm{GuardBlock}$ is part of a \textsf{Given} word, then the resulting guard restricts the values of location variables, thus, $\nonterm{VarID}$ words in it are not interpreted as gate parameters. Conversely, if the $\nonterm{GuardBlock}$ occurs in $\nonterm{When}$ or $\nonterm{Then}$, each $\nonterm{VarID}$ word is interpreted as a parameter unless the token \texttt{stored} is prepended to the $\nonterm{VarID}$ to mark it explicitly as a location variable.



We now introduce the concrete function definition for a \textsf{Guard} consisting of primitive, array, or structure variables.

\paragraph{Primitive Guard}
Words in $\nonterm{PrimGuard}_P$ define guards for variables with types in $T_p$. Consider a word $w_{\text{pg}} \in \nonterm{PrimGuard}_P$ of the form: 
\begin{smallmath}
\word{pg} = \text{\texttt{is }} \word{op} \; \word{rhs}
\end{smallmath}
with $\word{op} \in \nonterm{Op}_P$ and $\word{v} \in \nonterm{ExpValue}_P \cup \nonterm{VarRef}_P$. We define $\semmap[PrimGuard]{\word{pg}, x}$ which yields a predicate obtained by applying the operator $\semmap[Op]{\word{op}}$ to a left-hand variable $x$ and a right-hand operand.
The right-hand operand is either:
\begin{itemize}
    \item an expected, concrete value $\semmap[ExpValue]{\word{rhs}}$, if 
    $\word{v} \in \nonterm{ExpValue}_P$, or
    \item a variable reference, if $\word{rhs} \in \nonterm{VarRef}_P$. 
\end{itemize}
With the non-terminal $\nonterm{ExpValue}$ we denote a group of generic syntactic categories such as strings, natural or real numbers, ranges or arithmetic expressions. We assume their conventional interpretation and therefore omit further details. The interpretation of the variable reference, and the left-hand variable depends on the context. We distinguish two possible cases: a guard within a $\nonterm{Given}$ word, or within $\nonterm{When}$ or $\nonterm{Then}$ words. 

For a \textsf{PrimGuard} within a \textsf{Given} word, $\semmap[PrimGuard]{\word{pg}, x}$ yields a predicate over the location variable $\pi_{l}(x)$. If the right-hand side is a variable reference, i.e. $\word{rhs} \in \nonterm{VarRef}_P$, it is interpreted as a location variable $\pi_{l}(\semmap[VarID]{\word{rhs}})$.
For example, the word \texttt{VarA is greater or equal than 1} maps to the predicate $v_a \geq 1$ and the word \texttt{VarA is lower than VarB} maps to the predicate $v_a < v_b$, assuming $\pi_{l}(\semmap[VarID]{\text{VarA}}) = v_a$ and $\pi_{l}(\semmap[VarID]{\text{VarB}}) = v_b$.

For a \textsf{PrimGuard} within a \textsf{When} or \textsf{Then} word, $\semmap[PrimGuard]{\word{pg}, x}$ also yields a predicate, but over the switch parameter resulting from $\pi_{p}(x)$. If the right-hand side is a variable reference, i.e. $\word{rhs} \in \nonterm{VarRef}_P$, it is interpreted as a switch parameter unless it is preceded by the token \texttt{stored}, which means it is interpreted as a location variable. For example, the word \texttt{VarA is greater or equal than 1} now maps to the predicate $p_a \geq 1$. Furthermore, the word \texttt{VarA is lower than VarB} maps to the predicate $p_a < p_b$ and the word \texttt{VarA is equal to stored VarC} maps to the predicate $p_a = v_c$, where $\pi_{p}(\semmap[VarID]{\text{VarA}}) = p_a$, $\pi_{p}(\semmap[VarID]{\text{VarB}}) = p_b$ and $\pi_{l}(\semmap[VarID]{\text{VarC}}) = v_c$.


\paragraph{Array Guard}
Words derived from $\nonterm{ArrayGuard}$ define predicates for variables of type array by defining conditions that a subset of their elements must meet. Consider a word $\word{ag} \in \nonterm{ArrayGuard}_P$ of the form:
\begin{smallmath}
    \word{ag} = \text{\texttt{has }} \word{q}\ \word{n} \text{\texttt{ elements where each element }} \word{g} 
\end{smallmath}
with $\word{q} \in \nonterm{Quantifier}_P$, $\word{n} \in \nonterm{N}_P$ and $\word{g} \in \nonterm{Guard}_P$. 
We define the semantic mapping of such word, applied to a variable $x$ of type \texttt{array}, as $\semmap[Array Guard]{\word{ag}, x}$ which yields a predicate over a subset of elements of $x$. The subset is defined by the quantifier $\semmap[Quantifier]{\word{q}}$ (where $\word{q}$ may be an expression such as \texttt{at least}, \texttt{exactly}, \texttt{at most}) and a number of elements given by $\semmap[N]{\word{n}}$. The condition each of the elements $e$ in the subset must meet is given by $\semmap[X]{\word{g},e}$, where $\textsf{X}$ is the syntactic category of $\word{g}$: \textsf{PrimGuard}, \textsf{ArrayGuard} or \textsf{StructGuard}. For example, the word \texttt{VarA has at least 2 elements where each element has a value greater than 3} is mapped to the predicate $|\{ e \in v_a | e > 3\}| \geq 2$. Here, we assume that the word is a guard within a $\nonterm{Given}$, and $\pi_{l}(\semmap[VarID]{\text{VarA}}) = v_a$ is an array. 



\paragraph{Structure Guard}

Consider a word $w_{\text{sg}} \in \nonterm{Struct Guard}_P$ of the form: 
\begin{smallmath}
    \word{sg} = \text{\texttt{has attributes such that: }}\\ \text{\texttt{"}}\word[0]{at}\text{\texttt{"}} \word[0]{g} \word[0]{c} \dots \word[n-1]{c} \text{\texttt{"}}\word[n]{at}\text{\texttt{"}} \word[n]{g}
\end{smallmath}

The mapping $\semmap[Struct Guard]{\word{sg}, x}$, for variable $x$ of type \texttt{struct}, yields a predicate over the attribute values of $x$. This predicate is built from guards over attributes; the guard over the i-th attribute is given by $\llbracket \word[i]{g},\semmap[AttrID]{at_i} \rrbracket_{\mathsf{X}_i}$, where $\textsf{X}$ is the syntactic category of $\word[i]{g}$, \textsf{PrimGuard}, \textsf{ArrayGuard} or \textsf{StructGuard}; the concrete choice depends on the type of the attribute $\semmap[AttrID]{at_i}$. 
In guards of STS we may use helper function $\mathsf{get}_K : \texttt{struct}((k_0, \tau_0),\dots,(k_n, \tau_n)) \times \{k_0,\dots,k_n\} \to \valueuniv_{\tau_0} \cup \dots \cup \valueuniv_{\tau_n}$ to obtain the value of a given key in a structure.  
Conditions over attributes may be connected through conjunction operators, given by $\semmap[Conj]{\word[i]{conj}}$. For example, the word \texttt{"VarA" has attributes such that: "AttrA" has a value greater than 2 AND "AttrB" has a value equal to abc} maps to the predicate $\getValue[AttrA]{v_a} > 2 \land \getValue[AttrB]{v_a}$, assuming the word is a guard within a $\nonterm{Given}$, and $\pi_{l}(\semmap[VarID]{\text{VarA}}) = v_a$ is a structure with two attributes: an \texttt{integer} AttrA and a \texttt{string} AttrB. 


\begin{example}[Guard Block as part of a Given word]
    We consider again the example introduced in \autoref{lst:casestudy}, in particular, Scenario 1 (lines \ref{line:given-ini} to \ref{line:given-end}). Its precondition, preceded by \emph{\texttt{Given}}, determines the input guard of its corresponding STS, that is, $\STS_1$. As the word \emph{\texttt{the system is in its initial state}} is present in the precondition, then $\STS_1$ is considered a primary STS, that is, $\STS_1 \in \STSset_p$.
    The word \emph{\texttt{the user interface displays information on "enabledness", "availability" such that: "enabledness" has a value equal to true AND "availability" has a value equal to AV}} denotes that the input guard introduces restrictions over the variables \emph{enabledness} and \emph{availability}, connected by means of a logical \emph{and}.
    Considering the variable mapping introduced in \autoref{tab:variables}, this guard can be formally described as $IG^{1} \equiv (v_{\text{av}} = \text{AV}) \land v_{\text{en}}$.
\end{example}

\begin{example}[Guard Block as part of a When/Then word]
    We consider again our example introduced in \autoref{lst:casestudy}. As each scenario has exactly one input and one output switch, we denote their guards by $\phi_i$ and $\phi_o$, respectively. For readability, we define the predicate $\mathsf{insideCritSection}(d) \equiv (\getValue[l]{d} = v_\mathrm{cl}) \land (v_\mathrm{cs} <  \getValue[lp]{d} <  v_\mathrm{ce})$. This function evaluates to $\true$ if the detector, denoted with $d$, is inside the critical section (i.e. it has a matching lane and it is within the length position range), and evaluates to $\false$ otherwise. Then, using \autoref{tab:variables}, we define the guards for the input switches as follows:
    \begin{smallmath}
    \begin{array}{l}
    \phi_i^1 \equiv |\{ d \in p_\mathrm{fd} \mid \mathsf{insideCritSection}(d) \}| = 0 \land IG^1 \\[1pt]
    \phi_i^2 \equiv |\{ d \in p_\mathrm{fd} \mid \mathsf{insideCritSection}(d) \}| = 1 \land IG^2\\[1pt]
    \phi_i^3 \equiv |\{ d \in p_\mathrm{fd} \mid \mathsf{insideCritSection}(d) \}| \geq 2 \land IG^3\\[1pt]
    \phi_i^4 \equiv \true \land IG^4\\[1pt]
    \end{array}
    \end{smallmath}
    Guards $\phi_i^1$ to $\phi_i^3$ express the cases where zero, one, and two or more failing detectors are identified inside the critical road section. Simultaneously, the input guards $IG$ enforce that the system must be enabled, and the current report should be "Available". As the input in Scenario 4 has no parameters, no additional guard is present. 
    
    We can now define the guards of output switches:
    \begin{smallmath}
    \begin{array}{l}
    \phi_o^1 \equiv p_\mathrm{av} = \text{AV} \qquad
    \phi_o^2 \equiv p_\mathrm{av} = \text{PART AV} \\[1pt]
    \phi_o^3 \equiv p_\mathrm{av} = \text{NOT AV} \qquad
    \phi_o^4 \equiv \neg p_\mathrm{en}
    \end{array}
\end{smallmath}

Similarly, guards $\phi_o^1$ to $\phi_o^3$ represent the values that the system should report for its outputs: available, partially available or not available. Guard $\phi_o^4$ expresses the system is disabled. 
\end{example}

\subsubsection{Given}\label{subsubsec:Given}

The $\nonterm{Given}$ block states the preconditions of a $\nonterm{Scenario}$, as well as marking the initial state condition. Grammar \autoref{grammar:given} describes the $\nonterm{Given}$ production rules.

\begin{grammar}[h]
\caption{Pickles Grammar $P$: Given}\label{grammar:given}
\setlength{\jot}{1pt}
\vspace{-7pt}
\begin{smallmath}
\langle \text{Given} \rangle &::= \text{\texttt{Given}}~\langle \text{InitialCond} \rangle?~ \langle \text{Str} \rangle?~(\text{\texttt{"}}\langle \text{VarID} \rangle\text{\texttt{"}})\text{+}\\
&\phantom{:::=}\text{\texttt{such}}~\text{\texttt{that:}}~\langle \text{GuardBlock} \rangle\\
\langle \text{InitialCond} \rangle &::= \text{\texttt{The system is in its initial state}}
\end{smallmath}
\vspace{-7pt}
\end{grammar}

The presence of the \textsf{InitialCond} token determines if the corresponding scenario is in the subset $\STSset_p$. It is optionally followed by a string with no semantics; its sole purpose is to improve readability. The guard derived from the $\nonterm{GuardBlock}$ in $\nonterm{Given}$ determines the \emph{initial guard} of the system, denoted with $IG$, an additional condition over the location variables that the first switch must meet. Then, for some $\word{gi} \in \nonterm{Given}_P$, containing a word $\word{gb} \in \nonterm{GuardBlock}_P$, the resulting input guard $IG^i$ of the associated STS $\STS_i$ is defined as: $IG^i \equiv \semmap[Guard Block]{\word{gb}}$. 
If a scenario has no $\nonterm{Given}$ word, then $IG \equiv \true$.

\subsubsection{When and Then}\label{subsubsec:WhenThen}
The $\nonterm{When}$ non-terminal defines the interaction with the SUT through $\nonterm{InStep}$ words, while the $\nonterm{Then}$ specifies the observable response as a set of $\nonterm{OutStep}$. Grammar \autoref{grammar:when-then} defines the production rules for $\nonterm{When}$ and $\nonterm{Then}$.

\begin{grammar}[h]
\caption{Pickles Grammar $P$: When and Then}\label{grammar:when-then}
\setlength{\jot}{1pt}
\vspace{-7pt}
\begin{smallmath}
\langle \text{When} \rangle &::= \text{\texttt{When}}~\langle \text{InStep} \rangle~(\text{\texttt{And}}~\langle \text{InStep} \rangle)\text{*}\\
\langle \text{InStep} \rangle &::= \langle \text{InAction} \rangle~((\text{\texttt{"}}\langle \text{VarID} \rangle\text{\texttt{"}})\text{+}~\text{\texttt{such that:}}\\
&\phantom{:::=}~\langle \text{GuardBlock} \rangle)?\\
\langle \text{InAction} \rangle &::= \langle \text{Str} \rangle\\
\langle \text{Then} \rangle &::= \text{\texttt{Then}}~ \langle \text{OutStep} \rangle ~(\text{\texttt{And}}~ \langle \text{OutStep} \rangle)\text{*}\\
\langle \text{OutStep} \rangle &::= \langle \text{OutAction} \rangle ~((\text{\texttt{"}}\langle \text{VarID} \rangle\text{\texttt{"}})\text{+}~ \text{\texttt{such that:}}\\
&\phantom{:::=}~\langle \text{GuardBlock} \rangle)?\\
\langle \text{OutAction} \rangle &::= \langle \text{Str} \rangle
\end{smallmath}
\vspace{-7pt}
\end{grammar}

Let $\semmap[InAction]{\cdot} : \nonterm{InAction}_P \rightarrow \Gamma_i$ and $\semmap[OutAction]{\cdot} : \nonterm{OutAction}_P \rightarrow \Gamma_o$ map each input or output action word to the gate they represent. Again, this is a mere correspondence that we assume to be consistent and recorded, therefore we omit a constructive definition of these mappings. Let $\semmap[InStep]{\cdot} : \nonterm{InStep}_P \rightarrow \switchrel_I$ and $\semmap[OutStep]{\cdot} : \nonterm{OutStep}_P \rightarrow \switchrel_O$ map step words to their corresponding switch. 

Consider a scenario with
$m$ step words $\word{step} \in \nonterm{InStep} \cup \nonterm{OutStep}$, of the form:
\begin{smallmath}
    \word{step} = \word{a} "\word[0]{id}" ... "\word[q]{id}" \text{\texttt{such that:}} \word{gb}
\end{smallmath}
Thus, $\semmap[X]{\word{step}}$ for the $j$-th step is defined as follows:
\begin{smallmath}
\begin{aligned}
\semmap[X]{\word[j]{step}} &= (l_{j-1}, (\gamma_j, p_0, \dots, p_q), \phi_j, \psi_j, l_j), \quad 1 \le j \le m,\\[2pt]
\gamma_j &=
  \begin{cases}
    \semmap[InAction]{a_j}, &\text{if } \nonterm{X}=\nonterm{InStep}\\
    \semmap[OutAction]{a_j}, &\text{if } \nonterm{X}=\nonterm{OutStep}
  \end{cases}\\[2pt]
\phi_j &=
  \begin{cases}
    \semmap[GuardBlock]{g_j} \land IG, &\text{if } l_{j-1}=l_0,\\
    \semmap[GuardBlock]{g_j}, & \text{otherwise}
  \end{cases}\\[2pt]
(p_0,\dots,p_q) &= \pi_{p}(\semmap[VarID]{\word[0]{id}}) \dots, \pi_{p}(\semmap[VarID]{\word[q]{id}}),\\[2pt]
\psi_j &= \{\,\pi_{l}(\semmap[VarID]{\word[0]{id}}) := \pi_{p}(\semmap[VarID]{\word[q]{id}}) \mid 0 \le k \le q\,\}
\end{aligned}
\end{smallmath}

Given this transformation, each \emph{When} or \emph{Then} step becomes an input or output switch, respectively. These switches have \emph{actions} $\gamma$, defined by the textual description of the step, \emph{guards} $\phi$, given by guard blocks as introduced in \autoref{subsubsec:GuardBlock}, \emph{parameters} $(p_0,\dots,p_q)$ determined by the variable identifiers used in the step, and, finally, \emph{assignments} $\psi$; these store the value of the parameters in the corresponding location variable, with the intention of preserving this value for use in subsequent steps.

We assume that interactions are consistent in all steps across all scenarios, i.e. for any interactions  $(\gamma_1,\bar{p_1})$ and  $(\gamma_2,\bar{p_2})$ resulting from the translation, $\gamma_1 = \gamma_2$ implies that $\bar{p_1} = \bar{p_2}$. As a result, we can construct a set of interactions $\interact$, and also gates $\Gamma$, as the union of all interactions and gates, respectively, of the obtained switches resulting from the translation of all Pickles scenarios. We also assume that locations of scenarios are disjoint, i.e. for any pair of scenarios $\locset^1 \cap \locset^2 = \emptyset$. Note that it is easy to check these assumptions automatically.

To condense specifications, if a step's $\nonterm{GuardBlock}$ is a simple guard (i.e. without conjunction) over a primitive variable, we allow one-line expressions where $\nonterm{VarID}$ is immediately followed by $\nonterm{PrimGuard}$ (see \autoref{grammar:guard-block}), skipping the token \texttt{such that}; e.g., line \ref{line:synsugar} in \autoref{lst:casestudy}. This is merely a syntactic sugar and does not affect the translation.

\begin{example}
We can now define the partial models derived from our example introduced in \autoref{lst:casestudy}. As our specification suite has four scenarios, the resulting set of partial models is $\STSset = \{ \STS_1, \STS_2, \STS_3, \STS_4 \}$. We analyze one of them in further detail. \autoref{fig:sts-1} shows the STS resulting from Scenario 1. The gate denoted by \emph{faultyDetectors?} represent the input given by the word \emph{the controller detects "faulty detectors"} and \emph{display!} represents the output \emph{the user interface displays "availability"}.


\begin{figure}[b]
    \centering
    \scalebox{0.9}{
    \begin{tikzpicture}[->, >=stealth, auto, node distance=2cm,
        every state/.style={circle, draw, minimum size=0.7cm, inner sep=0pt, font=\small}, initial text=]
        
        \node[state, initial] (l01) {$l_0^1$};
        \node[state, below of=l01] (l11) {$l_1^1$};
        \node[state, below of=l11] (l21) {$l_2^1$};

        \path
        (l01) edge[dashed] node[right=0.45cm] {\small \shortstack[l]{faultyDetectors?($p_{\text{fd}}$)\\ 
        $|\{ d \in p_\mathrm{fd} \mid \mathsf{insideCritSection}(d) \}, =, 0) \land$\\ 
        $(v_{\text{av}} = \text{AV}) \land v_{\text{en}}$\\ 
        $v_{\text{fd}} := p_{\text{fd}}$}} (l11)
        (l11) edge[dashed] node[right=0.45cm] {\small \shortstack[l]{display!($p_{\text{av}}$)\\ 
        $p_\mathrm{av} = \text{AV}$\\ 
        $v_{\text{av}} := p_{\text{av}}$}} (l21);
    \end{tikzpicture}
    }
    \caption{$\text{STS}_1$: Faulty detectors outside the critical section.}
    \label{fig:sts-1}
\end{figure}
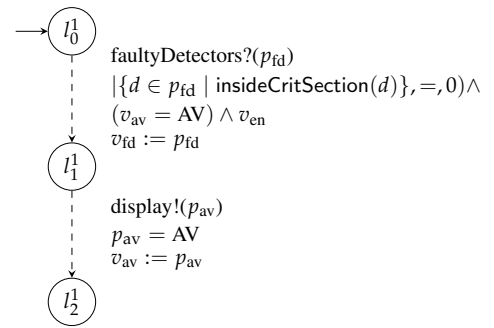
\end{example}

\section{Model Composition}\label{sec:modelcompos}

By applying the translation of \autoref{sec:grammar} on a specification suite in Pickles, we obtain a set of primary and secondary STSs, representing partial specifications of the system's behaviour. We are now interested in composing these partial specifications into a unified representation of the system's behaviour. 

To this end, we first introduce two composition operators that we apply to construct a master model. Choice composition ($\disj$) merges a set of STSs by unifying their initial location, enabling from it any transition that is initially enabled in one of the individual STS. Sequential composition ($\seqcomp$) merges the sink locations of a first STS with the initial location of the second. 

To construct the master model, we apply choice composition on both the primary STSs, and all STSs, yielding STS $\STS_{ini}$ and $\STS_{sys}$, respectively. Then, we sequentially compose $\STS_{sys}$ with itself, yielding $\STS_{sys} \seqcomp \STS_{sys}$, which denotes that a scenario may be followed by any scenario any number of times, if the guards of subsequent switches are satisfied. Lastly, we sequentially compose $\STS_{ini}$ with $\STS_{sys} \seqcomp \STS_{sys}$, yielding $\STS_{ini} \seqcomp \STS_{sys} \seqcomp \STS_{sys}$, to denote that an initial scenario must be executed before any other. The master model bundles alternatives between scenarios using choice composition, while sequential composition allows chaining multiple scenarios to generate all possible combinations.


\begin{definition}[STSs Choice composition]
Let $\STS_1 = (\locset^1, l^1_0, \vlset, \vpset,\Gamma,\interact,\switchrel^1)$ and $\STS_2 = (\locset^2, l^2_0, \vlset, \vpset,\Gamma,\interact,\switchrel^2)$ be two STSs.  
Their choice composition is the STS 
$\STS_1 \disj \STS_2 = (\locset^{\disj}, l_0^{\disj}, \vlset, \vpset, \Gamma,\interact, \switchrel^{\disj})$, where:
\vspace{-3.5pt}
\begin{smallmath}
\begin{aligned}
    \locset^{\disj}& = \locset^1 \cup \locset^2 \cup \{l_0^{\disj}\}\text{ so that } l_0^{\disj} \notin \locset^1 \cup \locset^2 \text{ is a fresh location}\\
    \switchrel^{\disj}& = (\switchrel^1 \backslash \{(l,\alpha,\phi,\psi,l') \in \switchrel^1 \mid l = l^1_0\})\\ 
    &\ \ \cup (\switchrel^2 \backslash \{(l,\alpha,\phi,\psi,l') \in \switchrel^2 \mid l = l^2_0\})\\
    &\ \ \cup\ \{ (l_0^{\disj} , \alpha, \phi , \psi, l) \mid (l_0, \alpha, \phi, \psi, l) \in \switchrel^1 \cup \switchrel^2 \wedge l_0 \in \{l_0^1, l_0^2\} \}\\
\end{aligned}
\end{smallmath}
For $n \in \mathbb{N}, n > 2$, we define $\bigdisj_{i \in [1,n]} \STS_i = \STS_1 \disj \STS_2 \disj \dots \STS_n$.
\end{definition}

\begin{definition}[STSs Sequential composition]
Let $\STS_1 = (\locset^1, l^1_0, \vlset, \vpset,\interact,$ $\Gamma,\switchrel^1)$ and $\STS_2 = (\locset^2, l^2_0, \vlset, \vpset,\interact,\Gamma,\switchrel^2)$ be two STSs, 
and let $\locset_s^1 = \{l \in \locset^1 \mid l\not\xrightarrow{\alpha,\phi,\psi} \}$ be the set of \emph{sink} locations of $\STS_1$, that is, the subset of locations without outgoing transitions. We define sequential composition 
$\STS_1 \seqcomp \STS_2 = (\locset^{\seqcomp}, l^1_0, \vlset, \vpset, \Gamma,\interact, \switchrel^{\seqcomp})$ where:
\begin{smallmath}
    \locset^{\seqcomp} &= \locset^1 \cup \locset^2 \backslash \{l_0^2\}\\
    \switchrel^{\seqcomp} &= \switchrel_1 \; \cup
    (\switchrel^2 \backslash \{(l,\alpha,\phi,\psi,l') \in \switchrel^2 \mid l = l^2_0\}) \; \cup \\ 
    &\{ (l^1, \alpha, \phi, \psi, l^2) \mid l^1 \in \locset_s^1, l_0^2 \xrightarrow{\alpha, \phi, \psi} l^2 \}
\end{smallmath}
\end{definition}


\begin{definition}[Master model of STSs]\label{def:synthesis-comp}
Let $\STSset_{p}$ and $\STSset_{s}$ be two disjoint sets of (primary and secondary) STSs, with $\STSset = \STSset_{p} \cup \STSset_{s}$. 
Then we define their \emph{master model} as $\STS_{ini} \seqcomp (\STS_{sys} \seqcomp \STS_{sys})$ where $\STS_{ini} = \bigdisj_{\STS_i \in \STSset_{p}} \STS_i$
and
$\STS_{sys} = \bigdisj_{\STS_i \in \STSset} \STS_i$.

\end{definition}

We note that a master model may be non-deterministic. It may also contain unreachable locations due to restrictions enforced by guards; these locations can be discarded after evaluating the satisfiability of the symbolic paths. Consequently, the master model may have unreachable scenarios; this provides feedback to the scenario writer that either the scenarios are wrong, or that such scenarios cannot be tested in the  system.

\begin{example}\label{ex-compos}
\autoref{fig:composed-STS} presents the master model $\STS_{master}$, obtained as a result of the synthesis operation $\bigdisj_{i \in [1,4]} \STS_i \seqcomp (\bigdisj_{i \in [1,4]} {\STS_i} \seqcomp \bigdisj_{i \in [1,4]} {\STS_i})$. In this example, we have omitted the paths that cannot be satisfied. For instance, no actions can be performed after executing the switches corresponding to Scenario 4, as after this, the system is no longer enabled, which is a precondition for all scenarios. We note that, in this example, all scenarios are marked as initial. This is because the specification was derived from BDD test cases, which were intended to be executed in isolation, from the initial state of the system, that satisfies some preconditions. 
\end{example}

\begin{figure*}[!h]
\scalebox{0.95}{
\begin{minipage}{0.7\textwidth}
    \centering
\composition
\end{minipage}
\hfill
\begin{minipage}{0.3\textwidth}
    \centering
    \footnotesize
    \begin{tabular}{p{0.75cm} p{3.5cm}}
        \hline
        Action & Definition\\
        \hline
        $i_1$? & \texttt{the controller detects "faulty detectors"} \\
        $i_2$? & \texttt{the controller access is lost} \\
        $o_1$! & \texttt{the interface displays "availability"} \\
        $o_2$! & \texttt{the interface reports status "enabledness"}\\
        \hline
    \end{tabular}
    
    \vspace{1em}
    
    \begin{tabular}{p{0.75cm} p{3.5cm}}
    \hline
    Guard & Definition\\
    \hline
    $\phi_i^1$ & $|\{ d \in p_\mathrm{fd} \mid \varphi(d) \}| = 0 \land (v_{\text{av}} = \text{AV}) \land v_{\text{en}}$ \\
    $\phi_i^2$ & $|\{ d \in p_\mathrm{fd} \mid \varphi(d) \}| = 1 \land (v_{\text{av}} = \text{AV}) \land v_{\text{en}}$\\
    $\phi_i^3$ & $|\{ d \in p_\mathrm{fd} \mid \varphi(d) \}| \geq 2 \land (v_{\text{av}} = \text{AV}) \land v_{\text{en}}$\\
    $\phi_i^4$ & $v_{\text{en}}$\\
    $\phi_o^1$ & $p_\mathrm{av} = \text{AV}$ \\
    $\phi_o^2$ & $p_\mathrm{av} = \text{PART AV}$ \\
    $\phi_o^3$ & $p_\mathrm{av} = \text{NOT AV}$ \\
    $\phi_o^4$ & $\neg p_\mathrm{en}$ \\
    \hline
    \end{tabular}

\end{minipage}
}
     \caption{$\STS_{master}$. Each line style (dashed, dotted, solid, dash-dotted) denotes transitions from $\STS_1$–$\STS_4$. Variables follow the reference in \autoref{tab:variables}.} 
    \label{fig:composed-STS}
\end{figure*}

\section{Test Generation and Translation to Pickles} \label{sec:test-cases}

Once a full representation of the SUT is obtained as a Symbolic Transition System applying the methods introduced in sections \ref{sec:grammar} and \ref{sec:modelcompos}, it is possible to obtain concrete test cases from it using model-based test generation techniques. We define a formal mapping to express the resulting test cases in Pickles syntax.

\subsection{Formal test cases} \label{sec:formaltest}

In this subsection we define formal test cases that can be generated from an STS. We adopt the generation methods proposed in \cite{covbasedtesting} to achieve 100\% switch coverage; therefore, we focus solely on the format of the resulting test cases and refer the reader to the original work for the underlying generation algorithms. Our formal test cases consist of a sequence of switches in the STS, initial values for location variables, and values for the switch parameters. In \cite{covbasedtesting} it is defined how such values can be found via a path condition. This is a Boolean formula that expresses under which conditions all guards along the path evaluate to true, considering the assignments that occur along the way. Values for the variables in the path condition that satisfy the formula can be determined by means of an SMT solver.

\begin{definition}[Formal test cases] \label{def:formal-test-cases}
Let $\STS$ be an STS.
    A \emph{formal test case} is a tuple $(r_0\dots r_n,\ini,\bar{w_0} \dots \bar{w_n})$ where:
    \begin{itemize}[label=--,itemsep=1pt, topsep=2pt, parsep=0pt, partopsep=1pt]
        \item $r_0\dots r_n$ is a sequence of switches, derived using techniques from \cite{covbasedtesting} 
        \item $\ini: \vlset \rightarrow \valueuniv$ assigns values to location variables
    \item $\bar{w_0} \dots \bar{w_n}$ is a sequence of value sequences,  such that the values satisfy the path condition for the switches $r_0\dots r_n$ as defined in \cite{covbasedtesting}.
    \end{itemize}
\end{definition}

\subsection{Translation to Pickles}

We now cover the mapping from formal test cases to Pickles syntax; in particular, we aim to generate executable test cases compatible with standard BDD tooling. While a formal test case defines values for the parameters of all switches, only those for input switches can be provided to the SUT; outputs are received from it, and may differ from those in the test case. Hence, in this translation, output values are ignored and replaced by the corresponding switch guards to validate the observed behaviour.

To define the translation from test cases to Pickles syntax, we introduce the notation $\invsemmap[\nonterm{X}]{\cdot}$; with it, we denote the mapping of mathematical elements to words of the sub-language $\nonterm{X}_P$ of the Pickles grammar $P$. If $\semmap[X]{\cdot}$ was introduced in \autoref{sec:grammar}, the definition of $\invsemmap[\nonterm{X}]{\cdot}$ is omitted as it is the inverse of $\semmap[X]{\cdot}$.

The syntax of a Pickles test case is displayed in Grammar \autoref{grammar:test-case}. We note that some non-terminals are shared with the specification suite (see Grammar \ref{grammar:overview}); they are reintroduced for clarity. 

\begin{grammar}[b]
\caption{Pickles Grammar $P$: Test Case}\label{grammar:test-case}
\setlength{\jot}{1pt}
\vspace{-7pt}
\begin{smallmath}
\langle \text{TestCase} \rangle &::= \langle \text{TestGiven} \rangle ~(\langle \text{TestWhen} \rangle ~|~ \langle \text{Then} \rangle)\text{*}\\
\langle \text{TestGiven} \rangle &::= \text{\texttt{Given the system}}\\
&\phantom{:::=}~\text{\texttt{is initialized with values:}}~ \langle \text{ValueDef} \rangle \text{+}\\
\langle \text{TestWhen} \rangle &::= \text{\texttt{When}}~ \langle \text{TestInStep} \rangle ~(\text{\texttt{And}}~ \langle \text{TestInStep} \rangle)\text{*} \\
\langle \text{TestInStep} \rangle &::= \langle \text{InAction} \rangle ~((\text{\texttt{"}}\langle \text{VarID} \rangle\text{\texttt{"}})\text{+} ~\text{\texttt{with values:}}\\
&\phantom{:::=}~ \langle \text{ValueDef} \rangle \text{+})?\\
\langle \text{Then} \rangle &::= \text{\texttt{Then}}~ \langle \text{OutStep} \rangle ~(\text{\texttt{And}}~ \langle \text{OutStep} \rangle)\text{*}\\
\langle \text{OutStep} \rangle &::= \langle \text{OutAction} \rangle ~((\text{\texttt{"}}\langle \text{VarID} \rangle\text{\texttt{"}})\text{+} ~\text{\texttt{such that:}}\\
&\phantom{:::=}~ \langle \text{GuardBlock} \rangle)?\\
\langle \text{ValueDef} \rangle &::= \text{\texttt{"}}\langle \text{VarID} \rangle\text{\texttt{"}} \text{\texttt{:}}~ \langle \text{Value} \rangle \\
\langle \text{Value} \rangle &::= \langle \text{N} \rangle ~|~\langle \text{R} \rangle ~|~ \langle \text{Str} \rangle ~|~ \langle \text{KeyValue} \rangle \text{+} ~|~ \langle \text{IndexValue} \rangle \text{+}\\
\langle \text{IndexValue} \rangle &::= \langle \text{N} \rangle \text{\texttt{:}}~ \langle \text{Value} \rangle\\
\langle \text{KeyValue} \rangle &::= \text{\texttt{"}}\langle \text{AttrID} \rangle\text{\texttt{"}} \text{\texttt{:}} \langle \text{Value} \rangle\\
\end{smallmath}
\vspace{-7pt}
\end{grammar}

Consider a formal test case $(r_0\dots r_n,\ini,\bar{w_0} \dots \bar{w_n})$, as defined in \autoref{def:formal-test-cases}. We first introduce an example of how such test case can be expressed in Pickles syntax. 

\vspace{-2pt}
\begin{example}\label{example-testtransl}
We consider the test case $(\bar{r},\ini,\bar{w_0} \dots \bar{w_n})$, derived from the model $\STS_{master}$ presented in \autoref{fig:composed-STS}. Below we define switch sequence $\bar{r}$, where subscripts indicate the source and target locations of the each switch. This test starts in the initial location $l_0$ and follows the trace up to location $l_{16}$. Initial values for location variables are defined by $\ini$; their reference can be found in \autoref{tab:variables}. Finally, $\bar{w_0} \dots \bar{w_n}$ denotes the parameter value for each switch where, again, subscripts indicate the source and target locations, and $\epsilon$ denotes an empty sequence.
\vspace{-6.5pt}
\begin{smallmath}
\begin{aligned}
&\bar{r} = r_{0,5}\: r_{5,8}\: r_{8,10}\: r_{10,16} \\[0.8pt]
&\ini(v_{\text{av}}) = \text{AV},\quad 
\ini(v_{\text{en}}) = \true,\\[0.8pt]
&\ini(v_{\text{cl}}) = 1,\quad
\ini(v_{\text{cs}}) = 2.0,\quad
\ini(v_{\text{ce}}) = 2.5 \\[0.8pt]
 &\ini(v_{\text{fd}}) = \{l: 1,\, \mathit{lp}: 1.5\}, \{l: 1,\, \mathit{lp}: 1.5\}, \{l: 1,\, \mathit{lp}: 1.5\}
\\[0.8pt]
&\bar{w}_{0,5} = \{l: 1,\, \mathit{lp}: 2.0\}, \{l: 1,\, \mathit{lp}: 2.8\}, \{l: 1,\, \mathit{lp}: 2.2\},\: \bar{w}_{8,10} = \epsilon
\end{aligned}
\end{smallmath}

Note that value sequences $\bar{w}$ for output switches are ignored, as we check instead that observed values from the SUT satisfy their guards. \autoref{lst:testcase} presents the test case in Pickles syntax.
\end{example}
\begin{lstlisting}[language=PicklesDSL, caption={Pickles test case executing scenarios 3 and 4 subsequently.},label=lst:testcase, numbers=left]
(@\textcolor{dkblue}{Given the system is initialized with values:}@)
    "availability": (@\textcolor{black}{AV}@)
    "enabledness": (@\textcolor{black}{true}@)
    "critical section lane": (@\textcolor{black}{1}@)
    "critical section start": (@\textcolor{black}{2.0}@)
    "critical section end": (@\textcolor{black}{2.5}@)
    "faulty detectors":(@\label{line:fd-start}@)
        (@\textcolor{black}{1: \{"lane": 1, "length position": 1.5\}}@)
        (@\textcolor{black}{2: \{"lane": 1, "length position": 1.5\}}@)
        (@\textcolor{black}{3: \{"lane": 1, "length position": 1.5\}}\label{line:fd-end}@)
When (@\textcolor{black}{the controller detects}@) "faulty detectors" (@\textcolor{dkblue}{with values:} \label{line:when-test-start}@)
    "faulty detectors":
        (@\textcolor{black}{1: \{"lane": 1, "length position": 2.0\}}@)
        (@\textcolor{black}{2: \{"lane": 1, "length position": 2.8\}}@)
        (@\textcolor{black}{3: \{"lane": 1, "length position": 2.2\}}\label{line:when-test-end}@)
Then (@\textcolor{black}{the user interface displays}@) "availability" equal to (@\textcolor{black}{NOT AV} \label{line:test-then1}@)
When (@\textcolor{black}{the controller access is lost}@)
Then (@\textcolor{black}{the user interface reports status}@) "enabledness" equal to (@\textcolor{black}{false}@)
\end{lstlisting}

We will now cover the translation details: first, constructing the words of \textsf{ValueDef} (i.e., representations of variables and their values), then, we address the construction of \textsf{TestGiven}, \textsf{TestWhen} and \textsf{Then} words, and finally the complete \textsf{TestCase}.

\paragraph{Value Definition}
In the following, we assume we have the mapping $\invsemmap[\nonterm{Value}]{\cdot} : \valueuniv \to \nonterm{Value}_P$. Each word in $\nonterm{ValueDef}$ is a concatenation of the textual representation of a variable $v$ with the one of its value $u$:
\vspace{-3.5pt}
\begin{smallmath}
    \invsemmap[\text{ValueDef}]{v, u} = \invsemmap[VarID]{v} \text{\texttt{:}} \invsemmap[Value]{u}
\end{smallmath}
\vspace{-4.5pt}
Concretely, for any $u \in \valueuniv$, $\invsemmap[Value]{u}$ is defined as follows:
\begin{itemize}
    \item If $\sortufun{u} \in T_p$, then $\invsemmap[Value]{u}$ yields simply the textual representation of the value. For example: $\invsemmap[Value]{3} =$ \texttt{3}.
    \item If $\sortufun{u} = \texttt{array}(\tau)$, then $\invsemmap[Value]{u}$ yields a list with values of each element of $u$. 
    \item Finally, if $\sortufun{u} = \texttt{struct}(K, T)$ then $\invsemmap[Value]{u}$ yields a list with the key-value pairs of $u$. 
\end{itemize}
\autoref{lst:testcase} provides an example, particularly in lines \ref{line:fd-start}-\ref{line:fd-end}: as faulty detectors is an array of structures, its value definition covers each of its three elements, with values for each attribute. 


\paragraph{Given} In the \textsf{TestGiven} word, in addition to the initial token, a new \textsf{ValueDef} word is created for each location variable $v \in \vlset$, as we need to state the initial values of each. We define $\invsemmap[TestGiven]{\cdot} : \vlset \times (\vlset \to \valueuniv) \to \nonterm{TestGiven}_P$ as a mapping from location variables and their initialization function to a word in $\nonterm{TestGiven}_P$. Then, $\invsemmap[TestGiven]{\vpset, \ini}$ is defined as:
\begin{smallmath}
     &\text{\texttt{Given the system is initialized}}\\ &\text{\texttt{with values:}} \: \{ \invsemmap[Value Def]{v, \ini(v)} \:|\: v \in \vlset \}
\end{smallmath}
\paragraph{When}
For each input switch $r_j \in \switchrel_I$ from the formal test case, 
a \textsf{TestInStep} within a \textsf{TestWhen} word is produced. A new \textsf{TestWhen} is created if the preceding switch $r_{j-1}$ is an output, otherwise, only a new \textsf{TestInStep} word is appended, preceded by the token \text{\texttt{And}}. The $\nonterm{InAction}$ is instantiated with the textual representation of the gate $\gamma_j$ of $r_j$, i.e., the one stored after parsing the specification suite (see \autoref{subsubsec:WhenThen}). For each gate parameter $p_k \in p_0 \dots p_m$, an entry of \textsf{ValueBlock} is created, with the parameter identifier and its value $u_k \in \bar{w}_j$.
Let $\invsemmap[TestInStep]{\cdot} : \switchrel_I \times \valueuniv^* \to \nonterm{TestInStep}_P$ 
 map an input switch and a corresponding value sequence to a word. 
For $r_j = (l_j,(\gamma_j,p_0\dots p_m),\phi_j,\psi_j,$ $l_j') \in \switchrel_I$ and a sequence of values $\bar{w} = u_0 \dots u_m$, we define $\invsemmap[TestInStep]{r_j, \bar{w}}$ as:
\begin{smallmath}
    &\invsemmap[InAction]{\gamma_j} \invsemmap[VarID]{p_0} \text{\texttt{,}} \dots \text{\texttt{,}} \invsemmap[VarID]{p_m} \text{ \text{\texttt{with values:}} } \\
    &\invsemmap[Value Def]{p_0, u_0} \dots \invsemmap[Value Def]{p_m, u_m}
\end{smallmath}
Parameter identifiers appear twice in each $\nonterm{TestInStep}$; as a step may reference many parameters, each value in a \textsf{ValueDef} word must be explicitly linked. An example of this mapping can be found in \autoref{lst:testcase}, particularly in lines \ref{line:when-test-start} to \ref{line:when-test-end}. 

\paragraph{Then}
If $r_j$ is an output switch, i.e. $r_j \in \switchrel_O$, an $\nonterm{OutStep}$ within a $\nonterm{Then}$ block is produced. If $r_{j-1} \in \switchrel_I$, a new $\nonterm{Then}$ block is created. Otherwise, the $\nonterm{OutStep}$ word is appended to the existing one, preceded by the token \texttt{And}. As noted earlier, output values $\bar{w}_j$ provided by the test case are ignored, as we aim to evaluate the SUT's output values instead. Thus, the textual representation of output switches is recreated as-is from the specification suite; for example, the Then step for Scenario 3 in \autoref{lst:casestudy} (line \ref{line:S3then}) matches the corresponding Then step in \autoref{lst:testcase} (line \ref{line:test-then1}). 


\paragraph{Test case}
 A word $\word{tc} \in \textsf{Test Case}$ can then be defined as the initial Given step defined by $\invsemmap[Given]{\vlset, ini}$, followed by a sequence of $n$ steps corresponding to switches. For the $i$-th switch, the associated step is prefixed with \texttt{And} if the switch $i-1$ has the same type (input or output); otherwise, it is prefixed with \texttt{When} for input switches and \texttt{Then} for output switches.


\section{Implementation} \label{sec:implementation}

To test our approach, we have implemented a proof-of-concept that: (i) given a set of specification scenarios in Pickles syntax, returns the master STS model and (ii) given this master model and a set of test cases derived from it, returns these tests in Pickles syntax. The implementation was developed in Python, using the Lark library\footnote{https://github.com/lark-parser/lark}, a parsing toolkit with support for ENBF grammar. 
The tool and the instructions to reproduce the results presented in this paper are publicly available \cite{artifact}.

Our tool operates in two modes: specification and test translation. In specification translation mode, it takes as input a plain-text file containing Pickles specifications, such as those introduced in \autoref{lst:casestudy}. It generates both the definitions of individual STSs and their composition, output in JSON format. In this prototype version of the tool, the master model includes all possible transitions, regardless of path satisfiability. In test translation mode, the tool takes as input an STS in JSON format (including textual descriptions of variables and actions) together with a set of test cases, also represented as JSON and derived from the STS. It outputs the corresponding test cases in plain text Pickles format, such as the one introduced in \autoref{lst:testcase}.


\section{Discussion}\label{sec:eval}

The Pickles approach offers several complementary advantages over traditional BDD-style specification and testing. In this section, we analyse three key characteristics of the framework: (i) human-readable artifacts with well-defined formal semantics, (ii) automatic scenario composition, and (iii) input/output parametrization. To illustrate the advantages related with each, we revisit the example introduced in \autoref{sec:case-study} and compare the execution of the four scenarios presented in \autoref{lst:casestudy} under two settings: a conventional BDD approach, where scenarios are executed in isolation with concrete values, and Pickles. Note that we do not evaluate the quality of the original test suite provided by Technolution; instead, we show that, with limited additional effort to generalize behaviour and adapt standard BDD scenarios to Pickles, one can obtain the benefits traditionally associated with MBT while maintaining human-readable artifacts.

\subsection{Human-readable artifacts with formal semantics}
First, by providing a formally defined, deterministic DSL, our framework enforces specifications with precise syntax and semantics suitable for automated testing. Nonetheless, as it is grounded in BDD-style constructs, it also preserves the familiarity and readability of textual specifications. In addition, by abstracting from individual examples to generalized behaviour, Pickles specifications condense information that is typically scattered across requirements, test cases, and auxiliary documents. This improves communication with stakeholders both during requirements definition and on achieved test results. 

Equally important is the bi-directional translation between structured natural language and formal models. Human-readable specifications in PicklesDSL constitute a common language across experts with varying technical backgrounds, facilitating precise and collaborative development. At the same time, human-readable tests improve explainability: in many critical systems, it is necessary to provide reports detailing what has been tested and the outcomes of these tests. The recipients of such reports often lack technical expertise, making test scripts in a programming language, or formal models insufficient. Moreover, readable tests create a feedback control loop, enabling the authors of the specifications to review the generated test cases and determine whether the specifications require adjustments.

As our language keeps the Gherkin structure, the resulting test cases can be automated with standard BDD tooling. The implementation of test adapters remains a manual task as in BDD; still, the number of keywords to be automated is always bounded by the number of keywords defined in the specification.

\subsection{Automatic scenario composition}

Pickles automatically combines individual scenarios into a unified behavioural model. For reactive systems, this yields a more realistic representation than isolated scenarios, as typically specified in BDD, as it captures longer sequences of actions. Such sequences can uncover bugs that emerge from the interaction between successive actions, that would otherwise remain hidden if functionalities were executed in isolation \cite{SeqCompos}. Moreover, scenario composition enables a \emph{shift-left} testing approach: integration, system, and even acceptance-level tests can be generated and executed at early stages of development, leading to earlier fault detection \cite{shiftlefttesting}. 

\begin{figure}[b!]
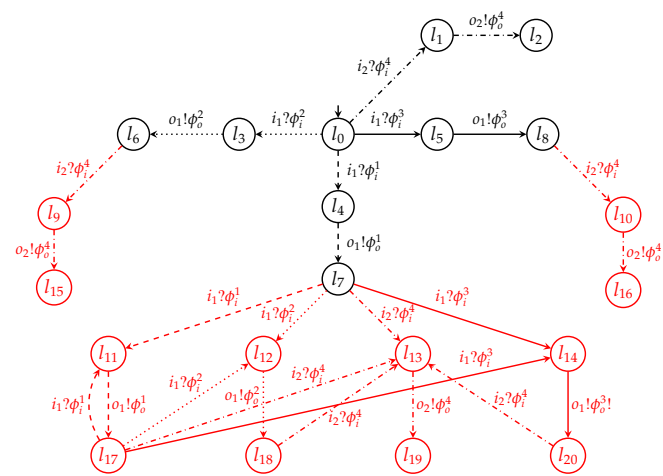

\centering
\resizebox{\columnwidth}{!}{%
\compositionCoverage
}
\caption{Coverage of $\text{STS}_{master}$. Transitions and states covered by only executing the specification scenarios are shown in black, whereas transitions in red show non-tested transitions.}
\label{fig:master-STS-cov}
\end{figure}

We illustrate now one of the key advantages of scenario composition: the improvement in transition coverage. Consider \autoref{fig:composed-STS}, showing the master model of the scenarios presented in  \autoref{lst:casestudy}. If the scenarios were executed in isolation, as is typical in BDD, they would cover only 8 of the 26 transitions (about 30\%). This limited coverage is visualized in \autoref{fig:master-STS-cov}, where covered transitions and states are denoted in black, and red represents non-tested transitions. In contrast, our framework, applying the techniques of \cite{covbasedtesting} to the same scenarios, yields test cases with 100\% transition coverage, an improvement of 70 percentage points. This gain does not imply manual definition of extra scenarios, however, it still needs some additional effort to generalize the system behaviour, including specifying variables and their domains.

From another perspective, consider again the model shown in \autoref{fig:composed-STS}. Achieving 100\% transition coverage through manually written test cases would require at least seven tests; the traces from $l_0$ to $l_{15}$, $l_0$ to $l_2$, $l_0$ to $l_{16}$ and the four possible traces from $l_0$ to $l_{19}$. Pickles achieves the same coverage providing only four scenarios as an input. This represents not only a reduction in manual effort, but also lower maintenance costs: when requirements change, fewer scenarios need to be updated.


\subsection{Input/output parametrization}\label{subsec:inoutparam}

Pickles supports the parametrization of inputs and outputs through explicit variable types and ranges. This enables early consistency checking at the specification level, as ill-typed expressions or incompatible ranges can be detected before any tests are executed. This feature is particularly useful when parts of the specification are produced by different teams, as interfaces can be checked early in the development process.

Moreover, moving from example-based scenarios to more general ones improves input coverage, as an arbitrary number of tests can be generated using any chosen heuristic. For example, consider, in \autoref{lst:casestudy}, the When step of Scenario 02 (see lines \ref{line:when-s2-start}-\ref{line:when-s2-end}).
The variable "faulty detectors", from now on referred to as $v_\text{fd}$, has as domain all arrays of length 1, 2, or 3 containing structures defined by a "lane" (an integer between 1 and 3) and a "length position" (a decimal between 1.0 and 3.0, exclusive). Suppose we generate tests as follows. First, we fix values for the start and end of the critical section to $v_{\text{cs}} = 1.5$ and $v_{\text{ce}} = 2.0$, and its lane to $1$, arguing (e.g. based on domain knowledge about our system) that varying these does not provide any gain. Applying boundary analysis, we choose four possible values for the detector's length position attribute: $1.001, 1.5, 2.0$ and $2.999$. 
With three lanes and four different length positions, there are 12 possible detector configurations.
Note that 2 of these configurations represent detectors in the critical section, i.e. those that satisfy the condition of lines \ref{line:cond-when-s2-start}-\ref{line:when-s2-end}, and 10 configurations don't.
%
We then have arrays of length 1, 2, or 3 containing
exactly one detector configuration in the critical section in 
and zero, one, or two detectors outside the critical section. 
%
The total number of valid arrays is then computed as follows, with combinations since order is irrelevant:
\vspace{-2pt}
\begin{smallmath}
\binom{2}{1}
\cdot
\left(
\binom{10}{0}
+
\binom{10}{1}
+
\binom{10}{2}
\right)
=
2\cdot(1 + 10 + 45)
=
112
\end{smallmath}

By using a single guarded input in the Pickles DSL combined with variable domain definitions, we represent 112 distinct input possibilities. This approach not only reduces the manual effort required to achieve high coverage but, again, also significantly improves test suite maintainability, as changes to the domain automatically propagate to all generated tests.

Extending this reasoning, the input switch in Scenario 01 allows for 175 possible values, while Scenario 03 allows for 11. Scenario 04, having no parameters, sees no benefit. If these scenarios were executed using only one concrete, hardcoded set of values, we would only be exercising approximately 1.5\% of the possible system inputs. In BDD, this coverage can be improved by adding more examples; still, this comes at the cost of manual effort and a larger, less maintainable test suite.

The benefits of generalizing the behaviour through input and output ranges are not limited to an improvement in coverage. This approach also reduces tester confirmation bias; when examples are written manually, testers tend to select input values that confirm their expectations about correct behaviour, rather than values that may expose errors \cite{ConfirmationBias}.

\section{Related Work}

The Pickles framework is conceived as a testing framework, but PicklesDSL, as shown throughout this work, is in itself a powerful formalism to unambiguously describe system requirements in a human readable form. Therefore, we review a wide set of related works 
on the following topics:
\begin{enumerate}[itemsep=1pt, topsep=2pt, parsep=0pt, partopsep=1pt]
    \item Natural-language support for requirement specification.
    \item Automated test case generation.
    \item Natural-language representation of test cases.
    \item Parametrization of requirements and test cases.
    \item Formal definition of requirements and test cases.
\end{enumerate}

The authors of \cite{IBDD} introduce an intermediate language that translates BDD scenarios into BDD Transition Systems (BDDTS), formal models suitable for automated test generation. This translation needs to be manually conducted; in this process, additional domain-specific information may be included in the transition systems. 
In comparison, Pickles condenses all the information needed for unambiguously modelling the system already in a natural language format, with no need for manual translation between readable and formal artifacts. Closely related, the work introduced in \cite{SeqCompos} defines a sequential composition operator to merge BDDTSs. This is an idea we adopt in our proposal, as part of the model composition introduced in \autoref{sec:modelcompos}.

Other authors have covered automatic test generation from structured natural language requirements. In \cite{CARVALHO2014275} test generation is automated from controlled natural language specifications in SysReq-CNL. Similarly, authors in \cite{RTCM} present RTCM, a restricted natural language for defining Test Case Scenarios from which concrete test cases are automatically derived. Both approaches support requirements parametrization, however, test cases are machine-readable only.

In \cite{OnlineMBTReusingModelsinIndustry}, model-based testing is performed using models generated by ComMA. This tool allows to define a component's behaviour and interfaces in a parametrized and compositional fashion. Similarly, TorXakis \cite{tretmansexistence2017} is an MBT tool with support for specifications with complex behaviour such as communication, synchronization, parallelism, non-determinism and data constraints. In both tools, however, neither the specifications nor the test cases are human-readable.

Other approaches generate and translate test cases from (semi-)formal models to human-readable form, i.e. comparable to the last two steps of our pipeline. An MBT pipeline that extracts human-readable test cases from UML diagrams is proposed by \cite{MBT-UML}. Additionally, \cite{UML-BDD} derives Gherkin-style acceptance criteria from UML specifications, generating human-readable tests. Still, both fail to incorporate parametrization, and are aimed specifically for UML users, as specifications are expected to be in this format.


Other work relies on structured natural language for defining formal requirements. Timed requirements and test cases can be expressed with the TEARS framework, presented in \cite{TEARS}. In \cite{propspecgrammar}, a method to specify properties for formal verification by means of Structured English Grammar is introduced. Moreover, authors in \cite{FRET} present FRET: a restricted, structured natural language to define requirements that are automatically translated to formally verified temporal-logic formulas. These approaches formally define parametrized system specifications in natural language, but do not provide test generation capabilities.



Finally, Natural Language Processing (NLP)-based testing techniques have grown, particularly with the raise of Large Language Models \cite{pradel2025regressiontestingnaturallanguage} \cite{LLMsforBDD} \cite{rao2025diffspecdifferentialtestingllms}. However, as noted by \cite{NatLangST}, many studies lack transparency in their evaluation, show narrow benchmarking, and their implementation details are often not disclosed. Still, NLP can play an important role, as noted by \cite{NLPforReqFormalization}: although it is not yet reliable for full automatic formalization from free natural language, it can support formal modelling when applied to structured requirements.

To the best of our knowledge, no existing approach simultaneously covers all topics listed at the beginning of this section. 

\section{Conclusions and Future Work}
This work introduces Pickles, a testing pipeline that bridges the gap between Behavior-Driven Development (BDD) and Model-Based Testing (MBT). By utilizing PicklesDSL, a Gherkin-based language for parametrized specifications and test cases, our approach enables non-technical stakeholders to collaborate on requirements that remain formally rigorous, and consistent. 
These specifications are automatically synthesized into a formal model, allowing for 
generating a wide range of test cases. 
These test cases are expressed back in our DSL, and can then be automated with standard BDD tooling.  

As shown through a case study with the company Technolution, using the same number of manually-defined scenarios, together with the additional, limited effort required to generalize the behaviour and migrate to the Pickles format, our framework can provide an increase of over 70\% in transition coverage and 98.5\% in input coverage compared to BDD. This result illustrates how we leverage MBT advantages while ensuring human-readable artifacts both for requirements and test cases.


For future work, we will expand our implementation and evaluate its performance across a diverse set of case studies, so that we validate our initial analysis of \autoref{sec:eval} at scale. In addition, as the framework's utility relies on its successful adoption by cross-functional teams, we will conduct empirical user studies to systematically evaluate the developer experience, while collecting feedback for further improvement of the tool.

Moreover, we aim to enrich our DSL with a wider range of expressions, e.g. control-flow sequences that allow non-deterministic behaviour. We also plan extensions that allow users to annotate scenarios with information about their criticality and probability of failure; this information can be integrated to the testing algorithm to help identify high-risk bugs faster. Given that timing is a primary constraint in critical systems, we also foresee incorporating time-constraint definitions.


Finally, we plan to investigate ways to assist the migration from BDD to our framework; in particular, how can pre-existing Gherkin test cases be processed, for example by leveraging large-language models, to automatically generalize the expected behaviour of the system from these examples.

\bibliography{bibliography}
\section*{About the authors}
\shortbio{Mar\'ia Bel\'en Rodr\'iguez}{is a PhD Candidate at the University of Twente (The Netherlands). \authorcontact[]{mariabelen.rodriguez@utwente.nl}}
\shortbio{Petra van den Bos}{is an Assistant Professor at the University of Twente (The Netherlands). \authorcontact[]{p.vandenbos@utwente.nl}}

\end{document}